\documentclass[twocolumn,letterpaper,aps,prc,longbibliography,superscriptaddress,showpacs,nofootinbib,floatfix,10pt]{revtex4-1}

\usepackage{graphicx}
\usepackage{xspace}
\usepackage{amsmath}
\usepackage{color}
\usepackage[%
  colorlinks=true,
  urlcolor=blue,
  linkcolor=blue,
  citecolor=blue
]{hyperref}

\usepackage{graphicx}
\graphicspath{{img/}}
\usepackage{dcolumn}
\usepackage{bm}
\usepackage{hyperref}

\usepackage{color}

\def\rlong     {R_{\rm long}}

\def\Rlong     {$\rlong$}

\newcommand{\mev}          {\ensuremath{\mathrm{MeV}}}

\newcommand{\gev}          {\ensuremath{\mathrm{GeV}}}
\newcommand{\gevc}         {\ensuremath{\mathrm{GeV}/c}}

\newcommand{\pp}           {\text{pp}}
\newcommand{\ppbar}        {\mbox{$\mathrm{p\overline{p}}$}}

\newcommand{\PbPb}         {\mbox{Pb--Pb}}
\newcommand{\pb} 		   {\mbox{Pb--Pb}}
\newcommand{\AuAu}         {\mbox{Au--Au}}
\newcommand{\pPb}          {\mbox{p--Pb}}

\newcommand{\pt}           {\ensuremath{p_{\mathrm{T}}}}

\newcommand{\pT}           {\ensuremath{p_{\mathrm{T}}}}
\newcommand{\mt}           {\ensuremath{m_{\mathrm{t}}}}
\newcommand{\snn}          {\ensuremath{\sqrt{s_{\mathrm{NN}}}}}
\newcommand{\sqrtsnn}[1]   {\ensuremath{\sqrt{s_{\mathrm{NN}}}=#1~\mathrm{TeV}}}
\newcommand{\sqrts}[1]     {\ensuremath{\sqrt{s}=#1~\mathrm{TeV}}}

\newcommand{\Npart}        {\ensuremath{N_\mathrm{part}}}

\newcommand{\Ncoll}        {\ensuremath{N_\mathrm{coll}}}

\newcommand{\abs}[1]       {\ensuremath{\left|#1\right|}}

\newcommand{\Figure}[1]    {Figure~\ref{#1}}

\newcommand{\com}[1]       {}

\newcommand{\rom}[1]{{\mathrm{#1}}}

\newcommand{\RAA}{\ensuremath{R_{\rom{AA}}}}
\newcommand{\RCP}{\ensuremath{R_{\rom{CP}}}}

\newcommand{\fm}           {\mathrm{fm}}

\newcommand{\cm}           {\mathrm{cm}}

\newcommand{\dedx}         {\ensuremath{{\rm d}E/{\rm d}x}}

\newcommand{\jpsi}         {\ensuremath{{\mathrm {J}}/\psi}}

\newcommand{\avbT}         {\ensuremath{\left< \beta_{\rm T}\right>}}

\newcommand{\Tfo}          {\ensuremath{{T}_{\rm kin}}}

\newcommand{\tn}[1]{\textnormal{#1}}

\newcommand{\kt}{\ensuremath{k_\tn{T}}}

\begin{document}

\preprint{APS/123-QED}

\title{Heavy-ion collisions - hot QCD in a lab}
\thanks{Lectures presented at the XIV International Workshop on Hadron Physics, Florianopolis, Brazil, March 2018.}%

\author{Mateusz P\l osko\'n}
 \altaffiliation[]{Lawrence Berkeley National Laboratory\\
 1 Cyclotron Road, Berkeley, 94720 California, USA}


\date{\today}

\begin{abstract}
High-energy heavy-ion collisions provide a unique opportunity to study the properties of the hot and dense strongly-interacting system composed of deconfined quarks and gluons -- the quark-gluon plasma (QGP) -- in laboratory conditions. The formation of a QGP is predicted by lattice QCD calculations as a crossover transition from hadronic matter (at zero baryochemical potential) and is expected to take place once the system temperature reaches values above 155 MeV and/or the energy density above $0.5~\gev/\fm^{3}$. The nature of such a strongly coupled QGP has been linked to the early Universe at some microseconds after the Big Bang. To characterize the physical properties of the short-lived matter (lifetime of about $10~\fm/c$) experimental studies at Relativistic Heavy-Ion Collider and the Large Hadron collider use auto-generated probes, such as high-energy partons created early in the hadronic collisions, thermally emitted photons, and a set of particle correlations that are sensitive to the collective expansion and the dynamics of the system. The lectures briefly introduced some of the experimental techniques and provided a glimpse at some of the results.
\end{abstract}

\maketitle



\section{\label{sec:introduction}Introduction}

Quantum Chromodynamics (QCD) solved on the lattice predicts a phase transition from normal nuclear matter to the state of deconfined quarks and gluons \cite{PhysRevD.85.054503}. The so-called quark-gluon plasma (QGP) is created once the temperature raises above a critical temperature $T_c$ of about 150 MeV at zero baryo-chemical potential, or density of the system is larger than about 0.5 GeV/fm$^3$.  Such conditions are often referred to as extreme: in units commonly used in daily life we would have to deal with densities of larger than $10^{15}~\mathrm{g}/\cm^{3}$ and temperatures beyond $10^{12}~\mathrm{K}$. Current understanding of the evolution of our Universe point out that such a QGP state could have existed microseconds after the Big Bang \cite{Schwarz:2003du}. Nowadays, matter under such extreme conditions can be created and studied in the laboratory by colliding heavy-nuclei at ultra-relativistic energies. Such collisions provide a unique opportunity for studying how the physical properties of the strongly interacting non-abelian partonic matter emerge from the fundamental interactions of QCD.

\begin{figure*}[tb]
	\centering
	\includegraphics[width=0.4\textwidth]{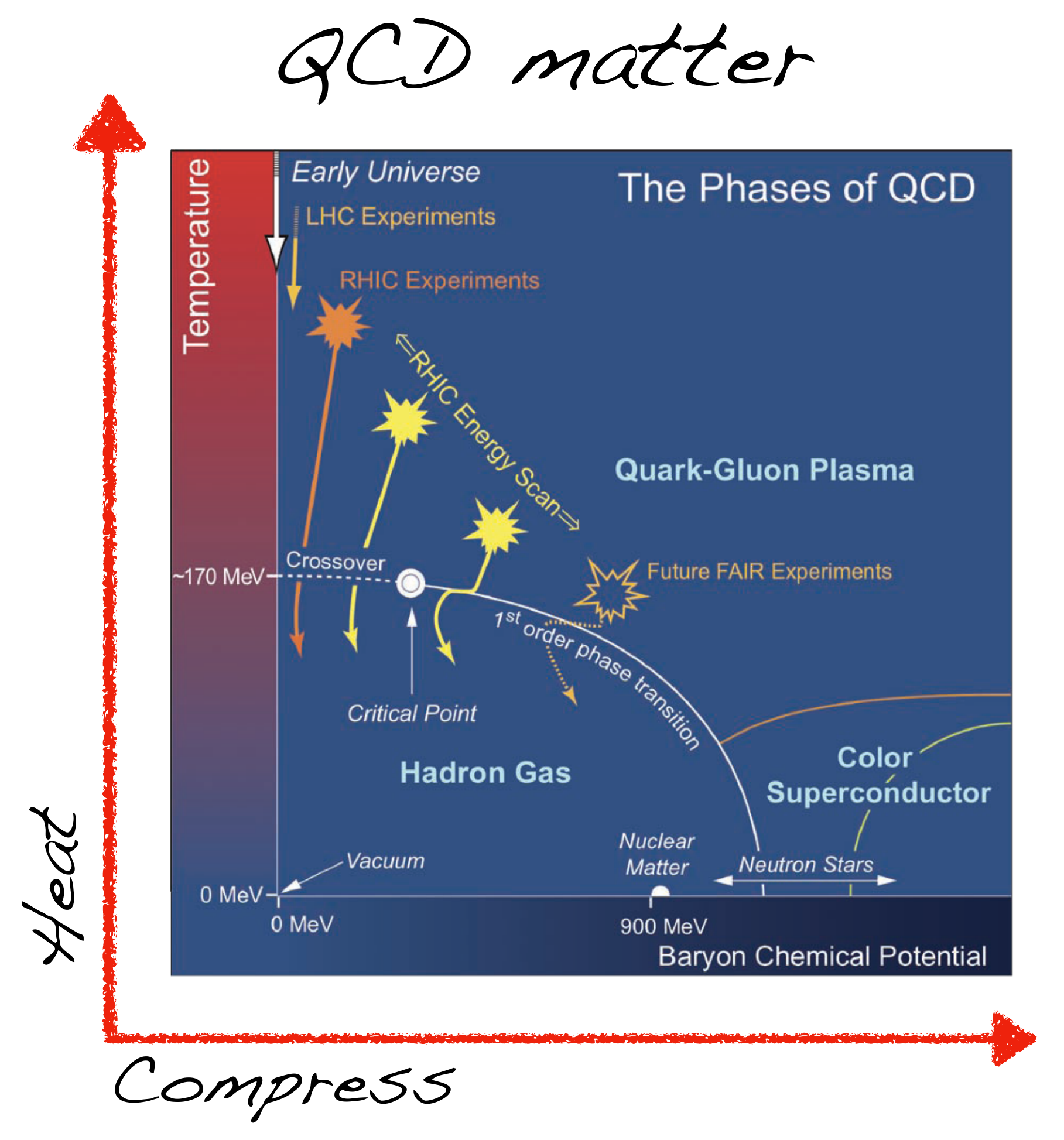}
	\includegraphics[width=0.42\textwidth]{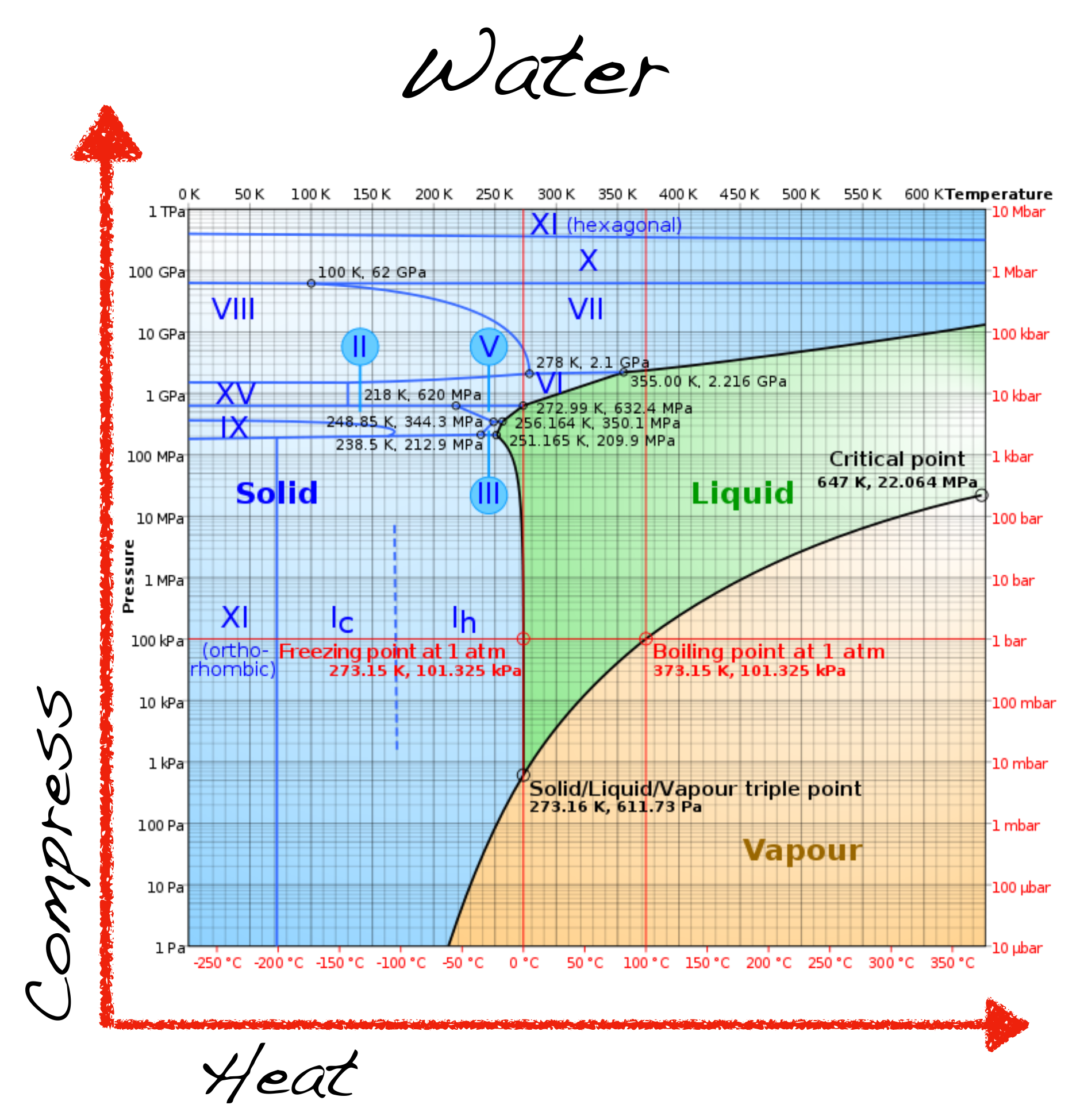}
	\caption{Phase diagram of QCD matter with a diagram for water. Several phases can be identified in both.}
	\label{fig:QCDdiagram}
\end{figure*}

There are a number of similarities between experimental studies of common materials and the studies QCD matter.
Figure \ref{fig:QCDdiagram} shows a sketch of a phase diagram of QCD matter and the similar temperature-density diagram for water.
Several well known phases of matter can be found in both diagrams: solid, gas, and liquid.
By compressing and/or heating the matter, different phases can be accessed experimentally.
High-energy heavy-ion collisions discussed in this write-up create matter at the high-temperature and low bayon density corner of the QCD diagram.

The experiments at top SPS and then RHIC energies brought the major discovery that the QGP behaves essentially as a strongly coupled liquid and is opaque to high energy partons (for example see Ref. \cite{Heinz:2000bk,Adams:2005dq} and references therein). These observations have been confirmed at the LHC \cite{Aamodt:2010pa,Aad:2010bu}; however, with an interesting twist that the collisions of much lighter system (proton-lead) can produce particle correlations that resemble observations from the heavy-ion collisions \cite{Abelev:2012ola} while providing no signal of in-medium modification of high-energy parton showers (jets) \cite{Acharya:2017okq}.

Model calculations show that by changing the collision energy one can access different regions of the QCD diagram.
High-energy heavy ion collisions have been studied experimentally in the last decades at increasing center-of-mass energies at the Brookhaven Alternating Gradient Synchrotron AGS ($\sqrt{s_\mathrm{NN}} < 5~\gev$), the CERN Super Proton Synchrotron SPS (up to \sqrtsnn{17.3}) and the Brookhaven Relativistic Heavy Ion Collider RHIC ($\sqrt{s_{\mathrm{NN}}} \leq 200~\gev$), and now also, at the Large Hadron Collider LHC at CERN at \sqrtsnn{2.76} and \sqrtsnn{5}, which is almost a factor 30 higher than the maximum collision energy at RHIC.
On the other hand, RHIC continues exploring the phase diagram in search of the critical point, with the Beam Energy Scan.

In this write-up I do not attempt to provide a complete set of results and experimental techniques employed in the heavy-ion experiments. I rather discuss a set of selected topics while providing an ample amount of references for further reading. I do focus on collisions at RHIC and at the LHC energies, which produce matter at the highest energy density in the largest volume and with the longest lifetime attainable in any laboratory experiment. The initial energy density achieved within heavy-ion collisions has been measured to exceed the critical energy density ($\epsilon_{c}$) for the phase transition to quark-gluon plasma (QGP). The analysis of current data combined with theoretical modeling indicates that the ultra-dense system of partons spends the first few $\mathrm{fm}/c$ after the collision above the $\epsilon_{c}$, which is much longer in comparison to typical strong interaction time scales, and may achieve a quasi/approximate-equilibrated state. However, the lifetime of QGP is far too short to study the system with external probes. Thus, characterization of the properties of the produced system must proceed by studying its decay products. A single heavy-ion collision is a complicated set of causally connected phenomena and physical processes. In order to infer the properties of QGP we must establish an understanding of the phenomena that are not related to the QGP phase but are a rather trivial consequence of colliding nuclei at high energies. For simplicity let us assume that a heavy-ion collision can be decomposed into three distinct stages arranged in time $(\tau)$:
\begin{itemize}
\item \textbf{Initial stage of the collision.} The initial stage comprises of the initial state radiation from the colliding nuclei and initial nucleon-nucleon inelastic collisions, including hard and soft parton-parton scatterings resulting in a large energy density within the overlap region. It is commonly agreed that the initial scatterings take place in the first tenths of fm/$c$ ($\tau \sim 0.3$ fm/$c$). Note, the actual passage time of the nuclei is $\tau_{pass}=2R/\gamma_{\rm cm}c$. Very often in order to interpret the experimental data it is important to know the impact parameter of the collision and the physical phenomena that control the initial energy density buildup. Such phenomena may involve saturation effects expected for low-x partons in nucleons and these effects are predicted much stronger for a nucleus as compared to single nucleons. Moreover, the initial random positions of the nucleons participating in the collision set the initial geometrical eccentricity of the colliding system. The understanding of the initial stage is critical for proper modeling and interpretation of experimental results obtained with heavy-ion collisions.
\item \textbf{Quark Gluon Plasma.} Owing to the many nucleon-nucleon collisions the energy density within the system can be sufficiently high such that the nucleons cease to exist as bound states of partons. The liberated quarks and gluons interact within a small but finite volume leading to the creation of a small droplet of a thermalized quark-gluon plasma. Within model calculations needed to interpret the experimental data it is usually assumed that the plasma reaches a point of thermal equilibrium at about $\tau \sim 1 $ fm/$c$. The main objects of the heavy-ion research are the physical properties of the QGP, such as temperature, density, shear viscosity / entropy density ratio. In this sense, the studies of QGP resemble the scientific program commonly known from the solid-state physics.
\item \textbf{Hadron gas and \textit{freeze-out}.} The expanding QGP cools down and freezes into color neutral hadrons at the so-called chemical freeze-out. After the relative abundances of particles are set elastic collisions between hadrons may still occur. Hadrons interact (only elastically) until the so-called kinetic freeze-out ($\tau > 10$ fm/$c$). The understanding of this stage of the collision is likewise critical for inferring the properties of QGP. In general, the interactions of particles in the hadron gas phase can distort the genuine QGP effects.
\end{itemize}

Lets introduce several experimental methods employed in studies of the properties of heavy-ion collisions that are used to infer the properties of the QGP:
\begin{itemize}
\item \textbf{Particle production and their \pt\ spectra.} The density of the produced particles contains information about the state of the hadron gas at freeze-out. Moreover, the relative abundance of the types of particles and their momentum distributions are affected by thermal properties of the created system, by the phenomena of the collective flow, and the stopping power within the collisions. It is worthwhile to note that because of the dynamical evolution of the nuclear effects within the initial stages of the collisions and the dynamical evolution of the thermalized plasma the measured distributions in AA collisions cannot be trivially extrapolated from the existing measurements of \pp\ collisions.
\item \textbf{Particle correlations.} The strongly interacting system results in correlations of particles in the final state. These correlations are sensitive to many properties of the system, such as the size of the emitting source (via quantum correlations of produced particles), the specific visosity - in particular, the shear viscosity / entropy density ratio (via the azimuthal momentum anisotropy of the produced particles), and at high-\pt, the jet-like correlations are sensitive to the high energy parton-medium interactions.
\item \textbf{Photons and di-lepton mass spectrum.} The low-momentum direct photons and low and intermediate mass di-leptons are expected to be sensitive to the dynamics of the hot fireball.
In particular, dileptons provide an experimental access to the thermal radiation of the QGP via the process $q\bar{q} \rightarrow \gamma^{*} \rightarrow e^+e^-$.
Experiments at SPS and RHIC have measured the excess in the di-lepton spectra consistent with the predictions of an enhanced thermal emission from the plasma.
Moreover, dileptons are perceived as a unique tool providing insights into the nature of the restoration of chiral symmetry of QCD.
From the measured direct photon spectra at low momentum we infer the initial temperature of the medium: the inverse slope parameter $T$ of an exponential fit to data at RHIC gives $T \approx 220$ \mev\ and now at the LHC $T \approx 300$ \mev. 
\item \textbf{Jets.} Jets of particles originated by highly virtual partons are used as auto-generated probes of the medium. The parton-medium interactions and subsequent modifications of the jet structure ({\it jet quenching}) are a sensitive to the density and temperature of the medium. The main goal here is to extract the so-called jet quenching coefficient $\hat{q}$ that is characteristic to a given medium. The coefficient is proportional the density of the medium.
\item \textbf{Heavy-quarks.} Heavy-quarks are one of the most promising tool to study the mass and flavor dependence of jet quenching, but also provide a stringent constraints on the transport properties of the medium such as the longitudinal drag - a diffusion coefficient.
\item \textbf{Quarkonia.} Production of bound $q\bar{q}$ states of heavy-quarks is sensitive to the temperature of the medium. The measurements of the various quarkonia states are an important source of information for modeling of the QGP.
\end{itemize}

This writeup does not discuss all of those points explicitly nor in depth. This write up should serve as an introduction to few selected ideas about experimental aspects of heavy-ion collisions and should not be regarded as a complete review.


\section{\label{sec:properties}Extracting Properties of Quark-Gluon Plasma}

\subsection{Heavy-ion Collisions: Centrality and Nuclear Effects}

\begin{figure*}[htb]
   \centering
   \includegraphics[width=0.5\textwidth]{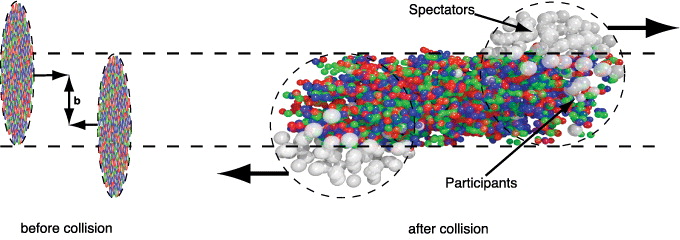} 
   \includegraphics[width=0.6\textwidth]{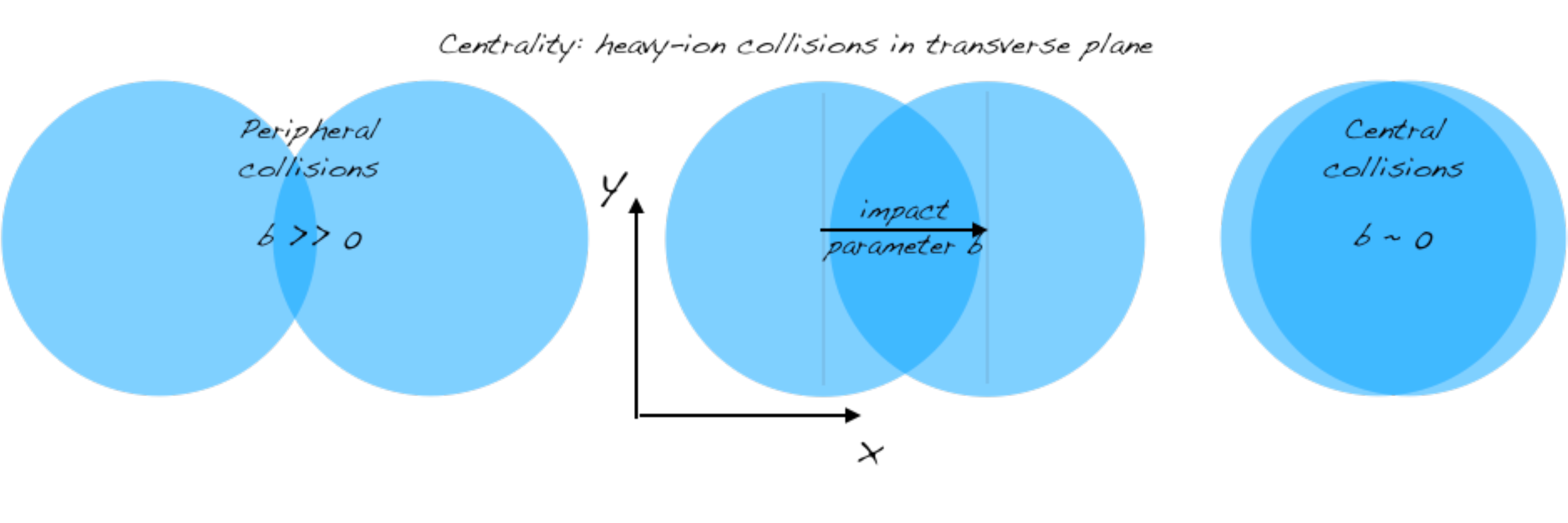}
   \caption{{\it Top Left:} Two heavy ions before the collision with the impact parameter $b$. {\it Top Right:} The spectators remain unaffected while in the participant zone, particle production takes place - UrQMD calculation \cite{Petersen:2008kb}. {\it Bottom:} An illustration of the often used nomenclature. Three collisions in the transverse plane are shown (from left to right): peripheral, semi-central, and central.}
   \label{fig:collision}
   \label{fig:centrality}
\end{figure*}

Before the quick overview of properties of QGP let us take a few notes concerning trivial consequences of colliding heavy nuclei at high energies.
First, by the virtue of colliding nuclei we expect that more than one nucleon may collide from each of the ions.
Also, some nucleons may encounter more than one nucleon on their flight path.
Figure \ref{fig:collision} shows two heavy ions before and after a collision - a visualization of UrQMD calculation \cite{Petersen:2008kb}.
The number of collisions and the number of nucleons that collided depends on the impact parameter $b$ of the collision (see Fig. \ref{fig:collision}).
A model of nuclear collisions connecting the geometry to the observed particle multiplicity was developed based on the inital work by Glauber \cite{Miller:2007ri}.
In a nutshell, from Glauber model we know how to connect the particle production to the number of colliding nucleons and their spacial position (geometry of the collision) -- the multiplicity is inversely proportional to the impact parameter of the collision.
Nowadays, heavy-ion experiments use multiple measures of multiplicity (or energy flow) within different regions of phase space to minimize uncertainties in determination of the total hadronic cross section and the simple geometry of nuclear collisions.
As a result the Glauber modelling of the nuclear reactions is rather well understood (including its limitations \cite{Acharya:2018njl}).

Apart from connecting multiplicity to the impact parameter the Glauber modelling is used to calculate other quantities characteristic of the collision geometry, such as the number of participating nucleons (\Npart{}), and the number of binary nucleon-nucleon collisions (\Ncoll).
Both of those quantities are often used to present the data.
\Figure{fig:centrality} shows sketches of few selected collision ``centralities'' in the transverse plane (perpendicular to beam axis) - peripheral collisions where impact parameter is large, semi-central, and central with $b$ close to zero.
Note, the impact parameter together with the axis along the beam define the so-called reaction plane.
Centrality of the heavy-ion collisions is often expressed in the percentile of the total inelastic cross-section.
Typically a peripheral collision is considered beyond 80\% of the cross-section (for example in Pb-Pb collisions) where the impact parameter approaches two times the radius of the nuclei.
While collisions referred to as ``central'' are at typically 0-10\% (or 0-5\%) of the cross-section.

It is worthwhile to note, that particle production can take place for collisions with $b>2R$ where it is dominated by electromagnetic processes.
These so-called ultra-peripheral collisions (UPC) are on their own an interesting field of research and among other interesting aspect provide unique insight into the structure of nuclei.
For a recent review of UPC see Ref.\cite{Klein:2017vua}.

Last but not least we should remark that the Lorentz contracted nuclei flying at each other with nearly speed of light are carriers of strong QCD fields.
Fields, that are of much larger magnitude as compared to single nucleons colliding in vaccuum.
Thus one of the complications in extracting the nuclear effects related to formation of QGP is the proper treatment of the so-called Cold Nuclear Effects (CNM).
The CNM effects such as modifications of the parton distribution functions within nucleons conained within the nuclei \cite{Eskola:2016oht} as compared to the free, unbound nucleons can affect interaction and production cross-sections.
Alternatively, the effects related to saturation of parton distribution functions at low-$x$ (fraction of momentum carried by partons) \cite{Mueller:2001fv} can also alter the particle production as compared to nucleon-nucleon collisions.
Moreover, calculations within the framework of the Color Glass Condensate Effective Field Theory \cite{Gelis:2010nm} show that non-trivial correlations between partons in the initial stages of the collisions may propagate to the final state hadrons \cite{Mace:2018yvl}.
These CNM effects are of crucial importance for accurate interpretation of the measurements in heavy-ion collisions.

\subsection{Particle production}

\begin{figure}[htb]
\begin{center}
\includegraphics[width=0.45\textwidth]{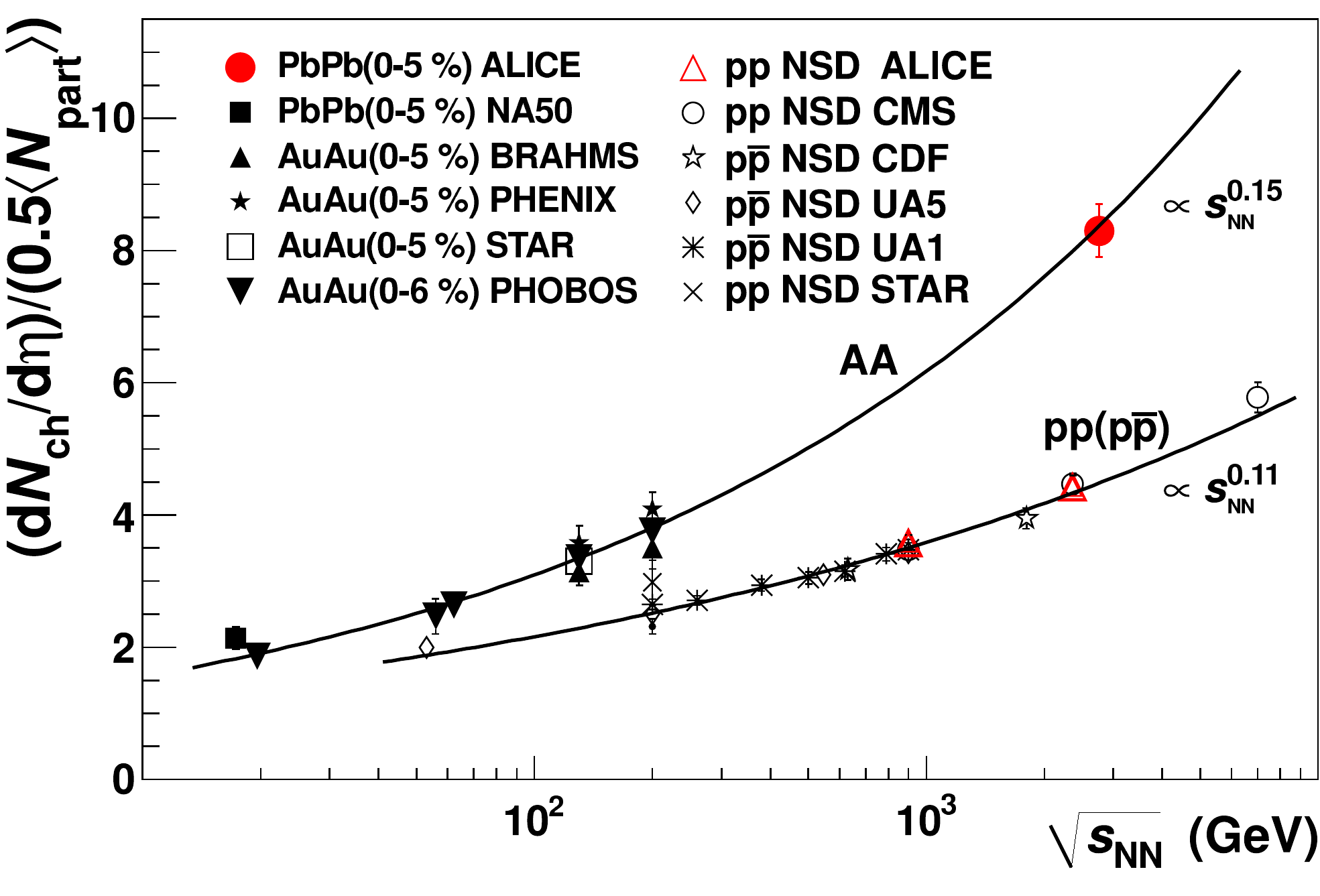}
\caption{{\label{fig3} Charged particle pseudo-rapidity density per participant pair for central nucleus--nucleus and non-single diffractive \pp~(\ppbar) collisions, as a function of $\snn$. The solid lines $\propto s_{\rm NN}^{0.15}$ and $\propto s_{\rm NN}^{0.11}$ are superimposed on the heavy-ion and \pp~(\ppbar) data, respectively.} Figure from \cite{Aamodt:2010pb}.}
\label{fig:dndeta_sqrts_Fig3.pdf}
\end{center}
\end{figure}

The interpretation of the experimental findings strongly relies on theoretical understanding and modeling of heavy-ion collisions. Any model providing sophisticated interpretations should be capable of reproducing some of the basic observables. One of those is the pseudo-rapidity density of the produced particles. Figure \ref{fig:dndeta_sqrts_Fig3.pdf} summarizes the experimental data on charged particle production in a narrow interval around the central rapidity ($\abs{\eta} \approx 0$) as a function of the collision energies in \pp\ and most central heavy-ion collisions. Within a single heavy-ion collision a single nucleon may interact more than once and, as expected, a heavy-ion collision produces on average larger particle density per participating nucleon than a \pp\ collision. Moreover, the increase of the energy of collisions for heavy-ions results in a much steeper rise than in the case of proton-proton collisions. An order of magnitude increase in the collision energy results in roughly doubling the energy density. At the top energy of the LHC the density of produced particles per participating nucleon pair in the most central collisions reaches a factor two times that in equivalent \pp\ collision. Note that the number of participating nucleons in a central \PbPb\ heavy-ion collision (0-5\%) is about 380, while for \pp\ collisions it is 2.

Observables based on particle identification play one of the central roles in the experimental studies of heavy-ion collisions.
The extraction of majority of signals of QGP depends on observables measured differentially with the type or the mass of the particle. The experiments employ various detection technologies to identify particles via their characteristic signals of their interactions with the detector material. These include detection of the Cherenkov radiation of an energetic particle traversing through a dielectric material of the detector, measurements of the energy loss within a gas or sensitive solid state volume of the detector, as well as measurements of the time-of-flight of particles. For an overview see the performance evaluation of several detection technologies employed in the ALICE detector \cite{Abelev:2014ffa}. More details on some high-energy heavy-ion collision detectors can be found in \cite{Aad:2008zzm} (ATLAS), \cite{Chatrchyan:2008aa} (CMS detector), \cite{Alves:2008zz} (LHCb detector), \cite{Adcox:2003zm} (PHENIX detector), and \cite{Ackermann:2002ad} (STAR detector).

\begin{figure*}[htb]
\begin{center}
\includegraphics[width=0.5\textwidth]{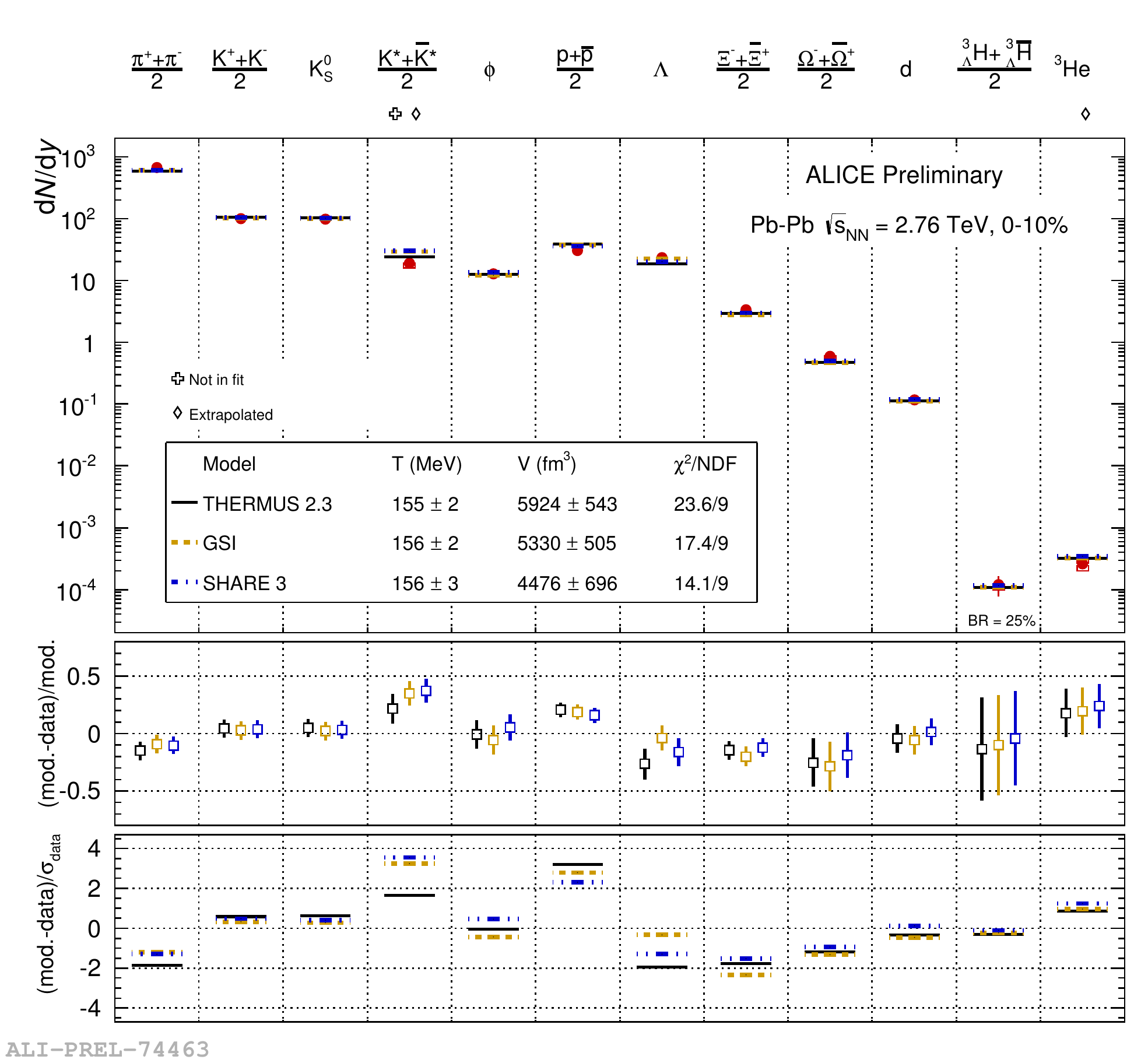}
\caption{Abundances of identified particles in most central heavy-ion collisions compared to thermal models GSI \cite{Andronic:2011yq}, THERMUS \cite{Cleymans:2006xj}, and SHARE \cite{Petran:2013lja}. Figure from \cite{Floris:2014pta}.}
\label{fig:thermalModels}
\end{center}
\end{figure*}

The so-called {\it thermal} or {\it statistical} models are formulated to describe the state of the system at the chemical freeze-out. Using the measured abundances of produced hadrons and assuming a local thermal equilibrium a model considering the grand canonical statistical ensemble is often used to extract the freeze-out temperature of the system. The grand canonical ensemble is defined by the chemical potential $\mu_{B}$, temperature $T$ an the volume $V$ of the system. In particular the number of hadrons produced in a collision can be written as:
\begin{align} 
n_{i}=g_{i}V\int\frac{\rm{d}^3p}{(2\pi)^3}\left[\exp{\left( \frac{E_{i}(p) - \mu_{i}}{T}\right)} - \epsilon\right]^{-1},
\end{align}
where the chemical potential $\mu_{i}$ for strongly interacting system is a linear function of the baryon chemical potential $\mu_{B}$, the strange chemical potential $\mu_{S}$, and the isospin chemical potential $\mu_{I}$. At the LHC the chemical potential is close to zero (at high collision energies as many anti-particles as particles are produced). The volume of the system is fixed by consideration of the \PbPb\ collision and it is mostly driven by the most abundant species (pions). Finally the temperature is predominantly constrained by the particle ratios with large mass differences (any baryon/pion ratio for example). Figure \ref{fig:thermalModels} shows an exemplary comparison of results obtained with models to the experimentally obtained particle abundances. Three models are shown from Refs. \cite{Andronic:2011yq,Cleymans:2006xj, Petran:2013lja}.
For a recent and an in-depth discussion of the thermal models and the data see (for example) Ref. \cite{Floris:2014pta} and references therein.

\subsection{System Size and Lifetime}

\begin{figure*}[htb]
  \centering
  \includegraphics[width=0.45\textwidth]{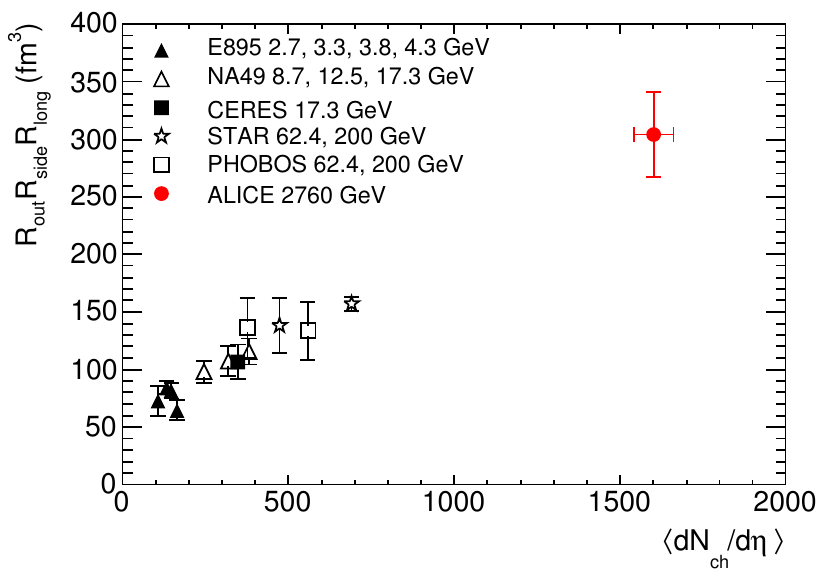}
  \includegraphics[width=0.45\textwidth]{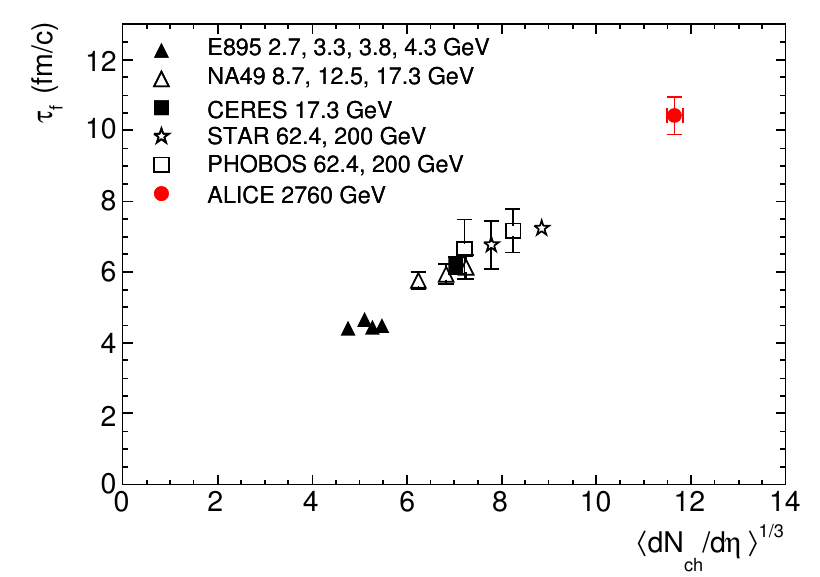}
  \caption{{\it Left: } Product of the three pion HBT radii at \kt~=~0.3\gevc. The ALICE result (red filled dot) is compared to those obtained for central gold and lead collisions at lower energies at the AGS, SPS, RHIC. Figure from Ref. \cite{Aamodt:2011mr} (see Refs. therein). \label{fig:hbt-1}
  {\it Right: } The decoupling time extracted from \Rlong(\kt). The ALICE result (red filled dot) is compared to those obtained for central gold and lead collisions at lower energies at the AGS, SPS, and RHIC. Figure from Ref. \cite{Aamodt:2011mr} (see Refs. therein). \label{fig:hbt-2}}
\end{figure*}

The size (the spatial extent) and the expansion rate of the system at decoupling from the hot plasma can be extracted by exploiting intensity interferometry, a technique which exploits the Bose-Einstein enhancement of identical bosons emitted close by in phasespace.
This approach, known as Hanbury Brown-Twiss analysis (HBT), was applied in $e^+e^−$, hadron-hadron, lepton-hadron, and heavy-ion collisions (see Ref. \cite{Aamodt:2011mr} and references therein).
Figure \ref{fig:hbt-1} (left panel) shows the product of three pion radii at a relative transverse momentum scale of the two pion extracted from central collisions at several energies.
Clearly the collisions at the LHC produce the largest spatial extent $R_\mathrm{out}R_\mathrm{side}R_\mathrm{long} \sim 300~\mathrm{fm}^3$.
Also, the decoupling time extracted using the longitudinal radii $R_{long}$ is largest for central collisions at the LHC reaching about 10~$\mathrm{fm}/c$ (see the right panel of Fig. \ref{fig:hbt-2}).

\subsection{Radial and transverse particle flow: kinetic freeze-out, and specific viscosity}
One of the consequences of the interactions within the expanding system can be velocity isotropization or thermalization of its constituents. In turn the bulk of the emitted hadrons will flow with a common average transverse velocity \avbT. Moreover, the bulk of the hadrons is likely to be emitted at the so-called freeze-out temperature \Tfo. The \pT\ spectra of the measured particles is be sensitive to these freeze-out parameters. Let us follow the prescriptions used in \cite{Abelev:2013vea}. The particle spectra are fitted individually with a blast-wave function~\cite{Schnedermann:1993ws}:

\begin{align} 
\frac{1}{\pt} \frac{\mathrm{d}N}{\mathrm{d}\pt} \propto \int_0^R r \mathrm{d}r\, m_{\rm T}\, I_0 \left( \frac{p_{\rm T}\sinh \rho}{T_{kin}} \right) K_1 \left( \frac{m_{\rm T}\cosh \rho}{T_{kin}} \right),
  \label{eq:blast-wave}
\end{align}

\noindent where the velocity profile $\rho$ is described by
\begin{align} 
 \rho = \tanh^{-1} \beta_{\rm T} = \tanh^{-1} \Biggl(\left(\frac{r}{R}\right)^{n} \beta_{s} \Biggr) \; .
 \label{eq:rhoBWdefintion}
\end{align}

\begin{figure}[htb]
\begin{center}
\includegraphics[width=8cm]{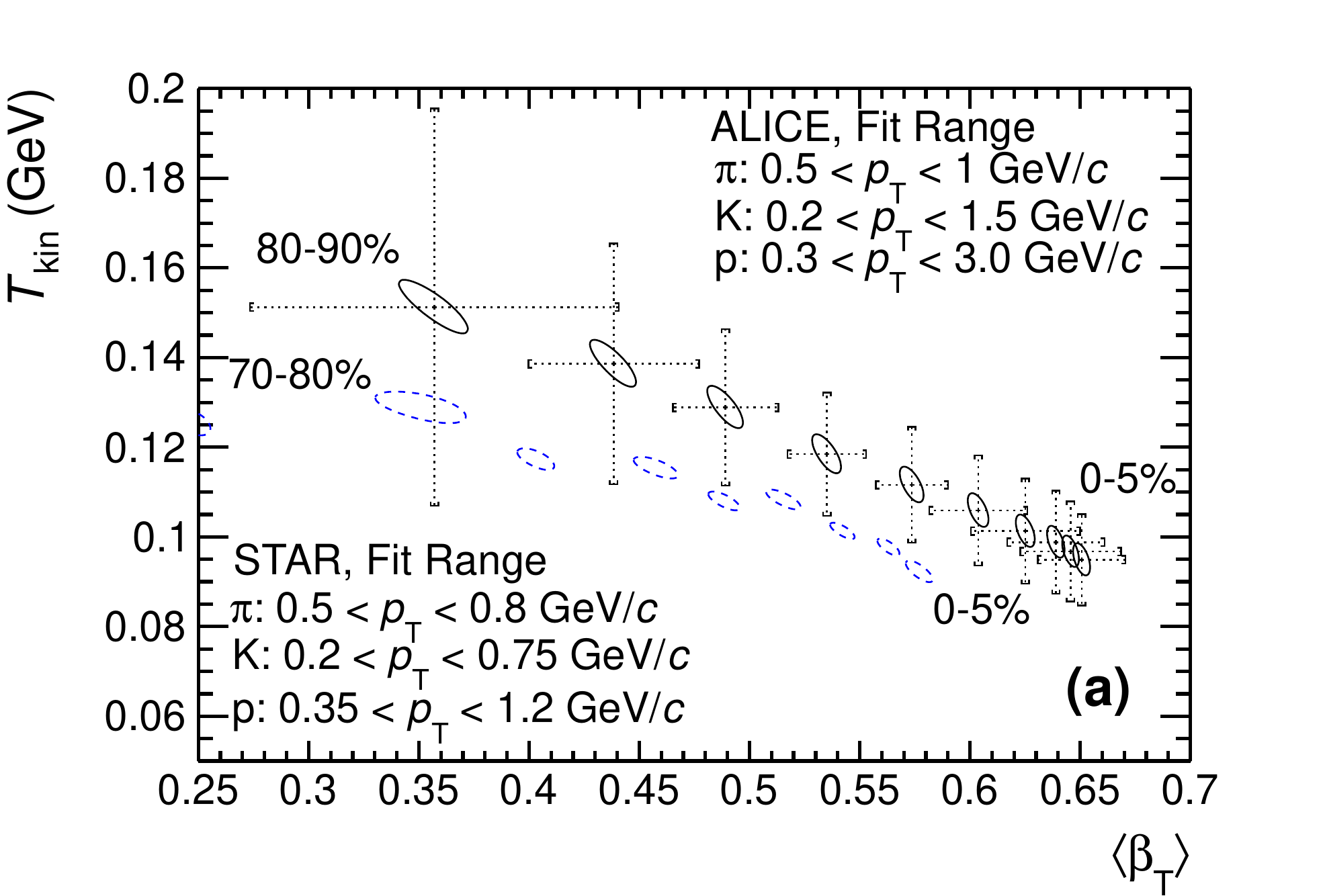}
\caption{Results of blast-wave fits at the LHC \cite{Abelev:2013vea}, compared to similar fits at RHIC energies~\cite{Adams:2005dq}; the uncertainty contours include the effect of the bin-by-bin systematic uncertainties, the dashed error bars represents the full systematic uncertainty (see text for details), the STAR contours include only statistical uncertainties. Figure from \cite{Abelev:2013vea}.}
\label{fig:cBlastWaveVsSTAR}
\end{center}
\end{figure}

The $\mt = \sqrt{\pt^2+m^2}$ is the transverse mass, $I_0$ and $K_1$ the modified Bessel functions, $r$ is the radial distance in the transverse plane, $R$ is the radius of the fireball, $\beta_{\rm T}$ is the transverse expansion velocity and $\beta_{s}$ is the transverse expansion velocity at the surface.
The free parameters in the fit are the freeze-out temperature \Tfo, the average transverse velocity \avbT\ and the exponent of the velocity profile $n$.
The function describes very well the spectrum of all particles within the measured \pt\ range.
However, from fits to a single particle species no physics interpretation of those parameters can be extracted.
The correct way to extract the freeze-out condition is to perform a combined fit to different particle species.
The results of the simultaneous fit defined by Eq. \ref{eq:blast-wave} to all the particle spectra within a selection of collision centralities are shown in Fig. \ref{fig:cBlastWaveVsSTAR} for ALICE measurements \cite{Abelev:2013vea} at the LHC and STAR at RHIC \cite{Adams:2005dq}.
The $T_{\rm{kin}}$ is lower for more central events as compared to peripheral events, suggesting a longer lived fireball.
This is strongly correlated with the evolution of the average expansion velocity - the \avbT\ grows for large volumes of the system (more central events).
Moreover, the almost an order of magnitude increase in the $\sqrt{s_\mathrm{NN}}$ results in the larger $T_{\rm{kin}}$ for all centralities at the LHC as well as it energizes the system to reach larger expansion velocities.
On the other hand, the similar trends of the blast wave parameters may also suggest a common physical properties of the system and its similar evolution despite the vastly different collision energies.
Note, that the radial flow (expansion of the system with common velocity of particles) provides a straightforward explanation for the so-called {\it the baryon anomaly} (see Ref. \cite{Abelev:2013xaa} and references therein for more details).
Finally, let us note that the kinetic freeze-out $T_{\rm{kin}}$ (about 100~\mev) is much smaller than the chemical freeze-out $T_{\rm{chem}}$ (about 155~\mev) extracted from the chemical abundances of particles.
This is of course consistent with the picture of the collision evolution where the chemical freeze-out is followed by the kinetic freeze-out.

\begin{figure*}[htb]
   \centering
   \includegraphics[width=0.6\textwidth]{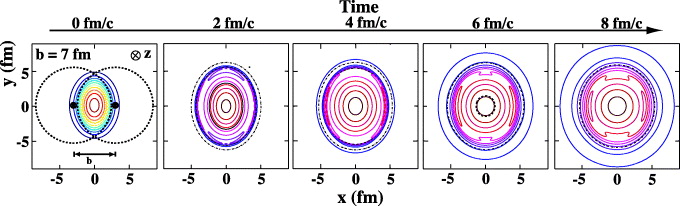} 
   \caption{Result of theoretical calculation \cite{KolbHeinz:EdensityVStime}. Time evolution of the initial transverse energy density profile in coordinate space for a non-central heavy-ion collision. The z-axis is along the colliding beams and the x-axis is defined by the impact parameter.}
   \label{fig:EdensityVStime}
\end{figure*}

\begin{figure}[htb]
\centering
\includegraphics[width=0.45\textwidth]{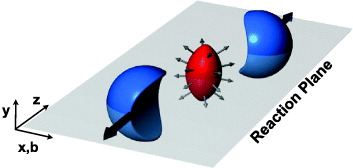} 
\caption{Sketch of two nuclei after the collision with the impact parameter $b>0$ creating the hot fireball in the center region. The reaction plane is defined by the $z$-coordinate and the vector of the impact parameter $b$ - the $x$-$z$ plane. The spatial anisotropy with respect to the reaction plane translates into a momentum anisotropy of the produced particles (anisotropic flow). Figure from \cite{1367-2630-13-5-055008}.}
\label{fig:reactionplane}
\end{figure}

One of the most interesting observations is the final-state azimuthal angular anisotropy of the produced particles in non-central collisions (with the impact parameter $b>0$) (see Ref. \cite{1367-2630-13-5-055008} for a good overview -- briefly followed below).
The measurements of these anisotropies allow to address fundamental questions on the dynamic properties of the hot and dense QCD matter.
An example of a non-central collision is shown in Fig. \ref{fig:reactionplane}.
The two colliding nuclei create an asymmetric fireball.
A correlation between the measured azimuthal momentum distribution of particles emitted from the decaying fireball and the initial spatial asymmetry is then driven by multiple interactions between the constituents of the created matter.
In other words, the momentum anisotropy depends on the initial spatial anisotropy but also carries information about how the matter flows.
The flow is directly related to the equation of state and thermodynamic transport properties of the created matter.
The initial asymmetry is quantified in the transverse plane by the so-called eccentricity
\begin{align} 
   \epsilon = \frac{\left<y^2\right>-\left<x^2\right>}{\left<y^2\right>+\left<x^2\right>},
\end{align}
where the $x$ and $y$ are the coordinates in the transverse plane of the collision. The interactions within the fireball may lead to the final energy-momentum anisotropy defined as:
\begin{align} 
   \epsilon_{T} = \frac{\left<T_{xx}-T_{yy}\right>}{\left<T_{xx}+T_{yy}\right>},
   \label{eq:epsilonT}
\end{align}
where $T_{xx}$ and $T_{yy}$ are the diagonal transverse components of the energy momentum tensor. An example of a model calculation demonstrating how the initial almond shaped fireball while cooling down expands in time is shown in Fig. \ref{fig:EdensityVStime}. The energy density asymmetries at late times can be seen by comparing the contours in the orthogonal $x$ and $y$ directions.

To characterize the various patterns of the particle distributions a Fourier expansion of the invariant cross-section is used:
\begin{align} 
   E\frac{{\rm d}^3N}{{\rm d}^3{\bf p}}=\frac{1}{2\pi}\frac{{\rm d}^2 N}{p_{\rm T}{\rm d}p_{\rm T}{\rm d}y}\left[ 1+2\sum_{n=1}^{\infty} v_n \cos{n(\varphi - \Psi_{RP}}) \right],
\end{align}
where $E$ and ${\bf p}$ are the energy and momentum of the particle, $\varphi$ is the azimuthal angle, and $y$ is the rapidity. For symmetry reasons the sine terms are zero, while the Fourier coefficients are given by: $v_{n}(\pT,y)=\langle \cos{n(\varphi-\Phi_{\rm RP})}\rangle$. Although the energy-momentum eccentricity $\epsilon_{T}$ from Eq. \ref{eq:epsilonT} is not directly observable one may experimentally study the evolution of the energy density by measuring the $v_n$ coefficients. To relate the $\epsilon_{T}$ to $v_n$ coefficients it is useful to note that the the elliptic flow coefficient $v_{2}$ can be written as $v_2= \frac{\left<p_{x}^{2}\right>-\left<p_{y}^{2}\right>}{\left<p_{x}^{2}\right>+\left<p_{y}^{2}\right>}$, where $p_x$ and $p_y$ denote the transverse components of the particle momentum.

\begin{figure}[htb]
\begin{center}
\includegraphics[width=0.45\textwidth]{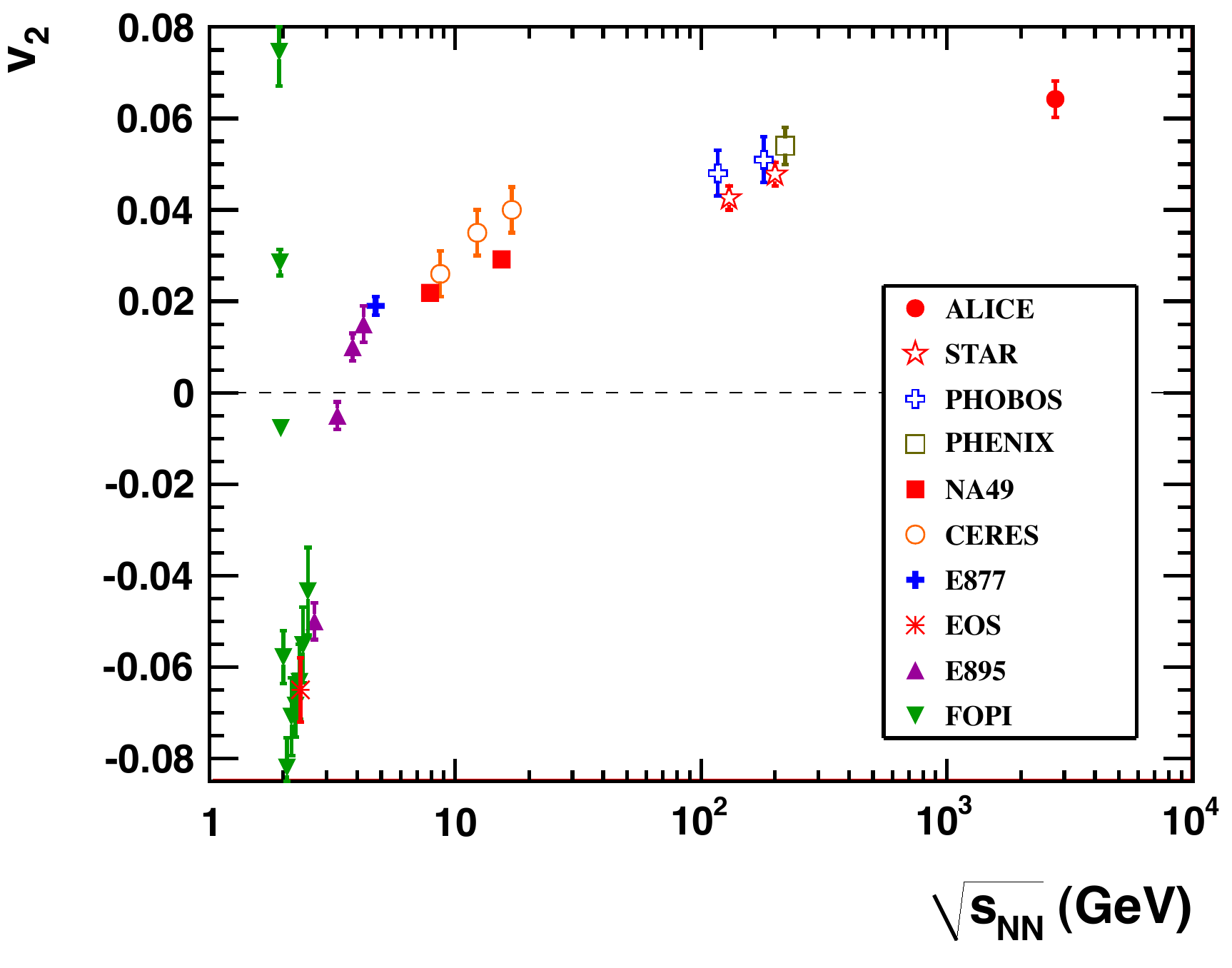}
\caption{Integrated elliptic flow at 2.76~TeV in \pb\ 20--30\% centrality class compared with results from lower energies taken at similar centralities~\cite{Voloshin:2008dg, Andronic:2004cp}. Figure from \cite{Aamodt:2010pa}.}
\label{1011.3914./arxiv.org/abs/1011.3914/img/plots/v2edep.pdf}
\end{center}
\end{figure}

Figure \ref{1011.3914./arxiv.org/abs/1011.3914/img/plots/v2edep.pdf} shows a compilation of integrated $v_2$ measurements in heavy-ion collisions as a function of the collision energy.
The figure prepared by the ALICE Collaboration \cite{ALICE:2011ab} demonstrates a continued increase of the magnitude of elliptic flow from few GeV, through RHIC energy to the LHC.
It is most interesting to see the transition of $v_2$ from the negative values (out-of-plane flow) at lower energies to a large positive (in-plane flow) values.
The negative $v_2$ at lower energies is reminiscent of strong nucleon potentials driving the produced particles to escape the dense fireball perpendicular to the reaction plane where the pressure is lower than in-plane.
On the other hand, the positive values at the lowest beam energies result from the in-plane rotational motion of the nucleons participating in the collision which predominantly happens in-plane.
Again the positive $v_2$ at energies $\sqrt{s_\mathrm{NN}} > 4$ GeV indicate an onset of strong interactions within the fireball that eventually at high energies lead to creation of a strongly coupled de-confined medium.
The results on $v_2$ from RHIC have established that for such a system, where the mean free path of a particle is much smaller than the size of the system, the evolution of QGP can be describbed by a relativistic hydrodynamics with a very small specific viscosity.
For more details on the application of hydrodynamics to heavy-ion collisions see \cite{Ollitrault:2008zz}.

\begin{figure}[htb]
\begin{center}
\includegraphics[width=0.45\textwidth]{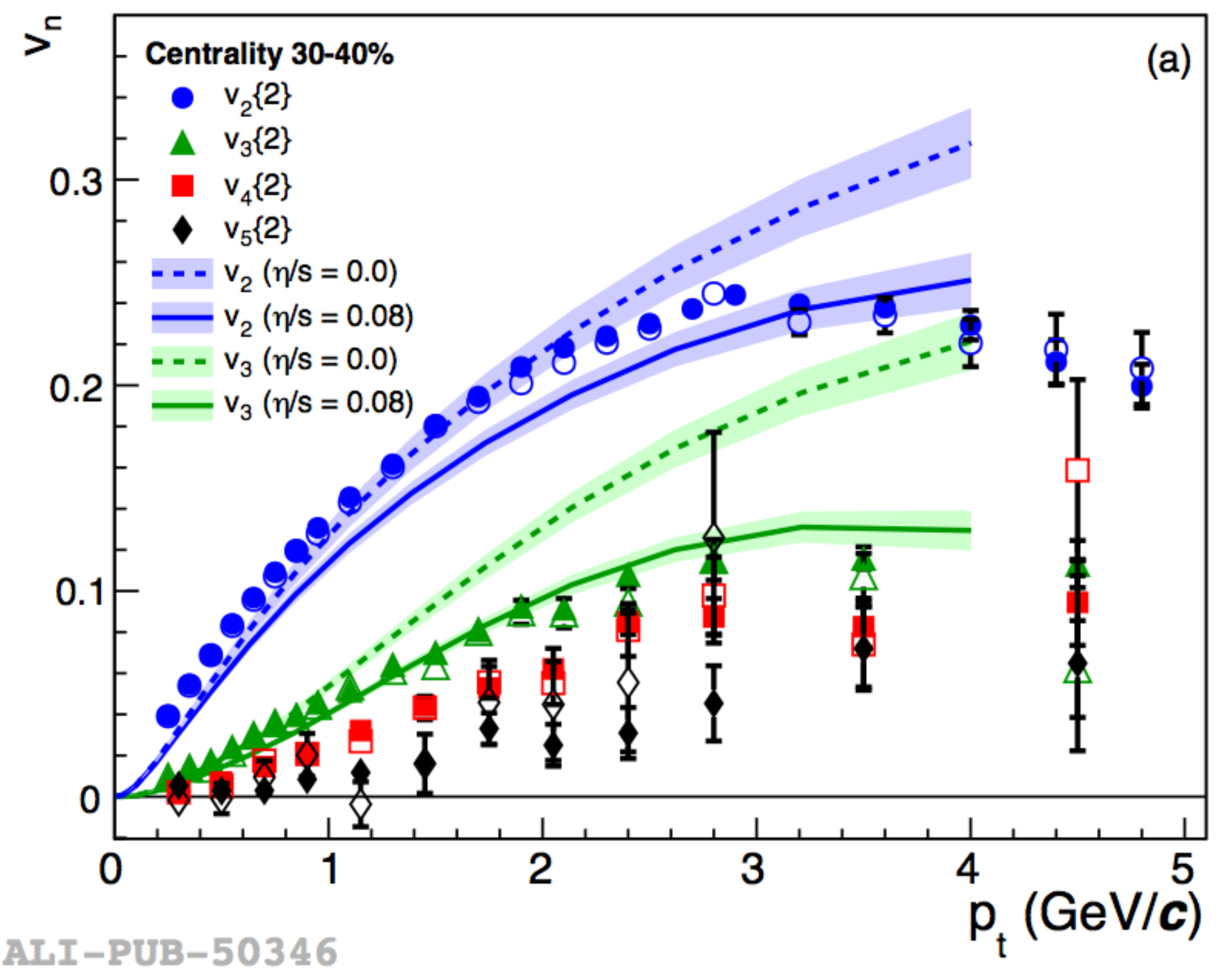}
\caption{{ $v_2$, $v_3$, $v_4$, $v_5$ as a function of transverse momentum and for three event centralities. The full, open symbols are for $\Delta\eta > 0.2$ and $\Delta\eta > 1.0$, respectively. The data in the 30\%--40\% most central collisions compared to hydrodynamic model calculations. } Figure from \cite{ALICE:2011ab}.}
\label{fig:vn_hydro_Fig3.pdf}
\end{center}
\end{figure}

Figure \ref{fig:vn_hydro_Fig3.pdf} presents the Fourier coefficients measured at the LHC in semi-central (30-40\%) collisions \cite{ALICE:2011ab}.
The $v_2$ is in particular sensitive to the internal friction or specific viscosity of the fluid, or more precisely, $\eta/s$, the ratio of the shear viscosity ($\eta$) to the entropy density ($s$) of the system.
Comparison of the elliptic flow measured in heavy-ion collisions with theoretical models suggests that the hot matter created in the collision flows like a fluid with little friction, with $\eta/s$ close to a limit derived using string theory methods with the anti-de Sitter / Conformal Field Theory conjecture \cite{Kovtun:2004de} for a perfect fluid $\eta/s = \hbar/4\pi k_{B}$, where $\hbar$ is Planck's constant and $k_{B}$ is the Boltzmann constant.

Further constraints to the theoretical calculations can be achieved by measuring the higher $n$ Fourier coefficients $v_n$ for $n>2$.
The odd coefficients arise from the event-by-event fluctuating orientations of the nucleons.
In some events the overlapping collision zone may be shaped similar to a triangle.
The $v_3$ is associated to the so-called {\it triangular flow}.
Further symmetry planes introduced by the geometrical fluctuations may result in large number of $v_n$ coefficients to be non-zero and can be used to further constrain the theory (see Fig. \ref{fig:vn_hydro_Fig3.pdf}).

\subsection{Temperature, Density and Transport Coefficient}

\subsubsection{Temperature}

Measurements of real and virtual photons can be used to characterize the temperature of the created medium.
For example, lepton pairs, are produced during the entire space-time evolution of the fireball and freely escape, undisturbed by final state interactions.
The virtual photons decaying into lepton pairs can be characterized by two variables, mass $M$ and transverse momentum \pt.
The mass distribution can be directly connected to the space-time averaged spectral function of the intermediate vector meson.
The measurement of \pt\ spectra of lepton pairs may offer access to their emission region, as \pt\ encodes the fireball temperature and the transverse flow.
For an overview and discussion of thermal dilepton production in heavy-ion collisions at SPS and RHIC see Ref. \cite{Rapp:2014hha} and references there in.
In constrast to hadrons, that flow with the characteristic velocities at decoupling, the lepton pairs are continuously emitted during the evolution of the system.
They are thus produced with small flow and high temperature at early times, and larger flow and smaller temperatures at later times.
With the measurement of direct photon spectrum in central heavy-ion collisions and in particular its thermal contribution the current understanding is that one of the hottest places in the Universe is created at the LHC.
The spectrum in $0.9<\pT<2.1~\gevc$ and 0-20\% centrality class can be described by an exponential with an inverse slope parameter of about 300~\mev\ \cite{Adam:2015lda}.
For a recent overview of electromagnetic probes of QGP see Ref. \cite{Paquet:2017wji} and references therein.

\begin{figure}[htb]
\begin{center}
\includegraphics[width=0.35\textwidth]{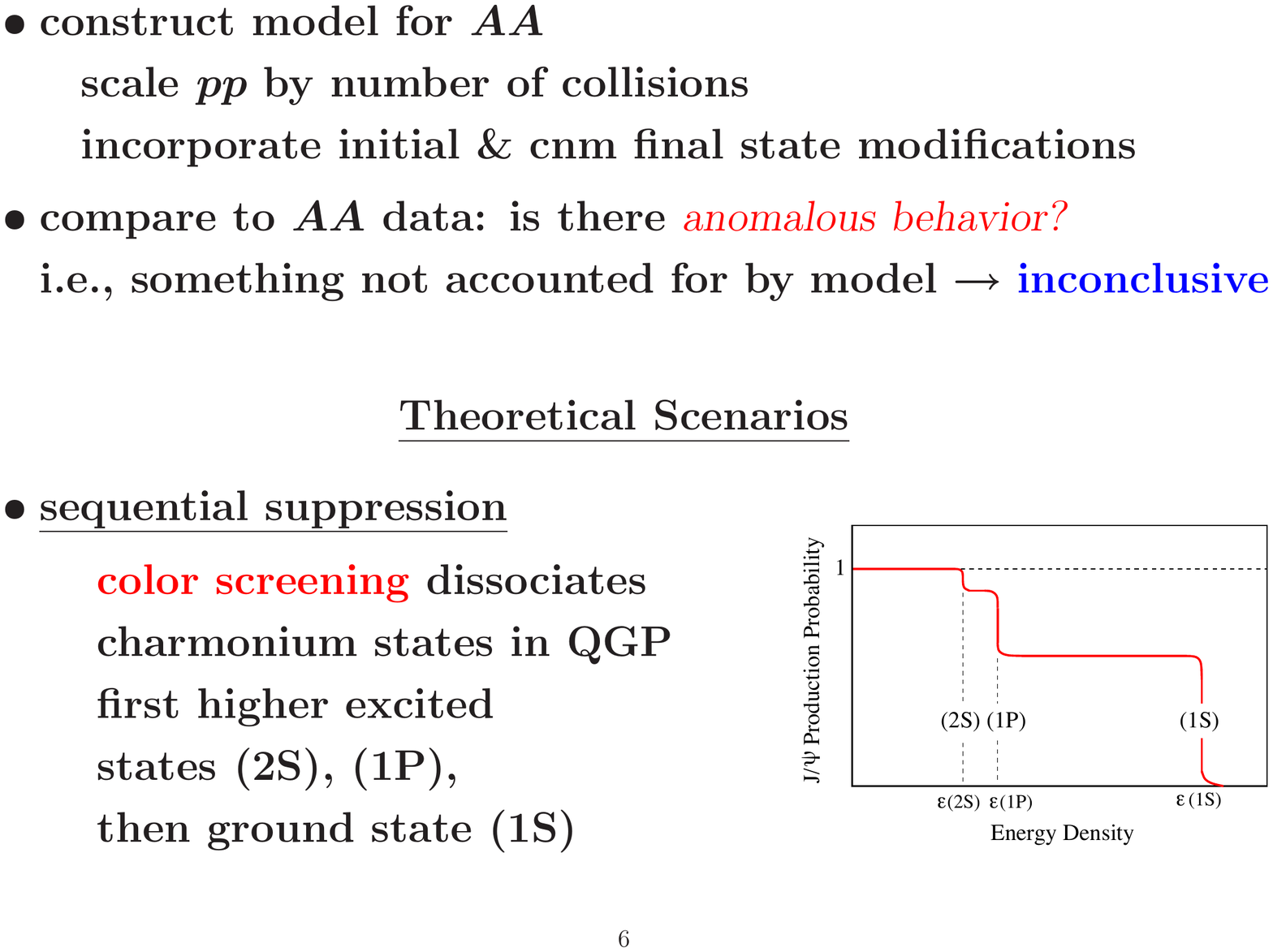}
\includegraphics[width=0.35\textwidth]{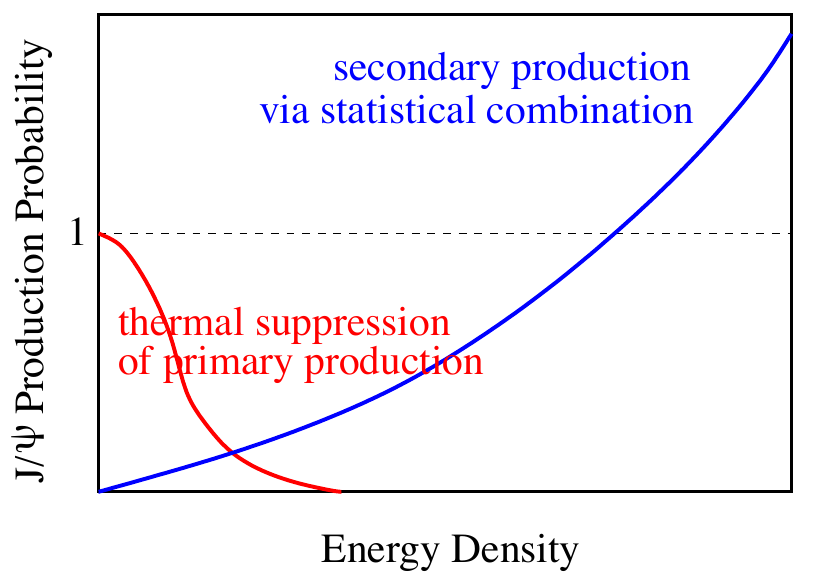}
\caption{{\it Top:} A schematic representation of suppression of charmonia states as a function of energy density. The probability is normalized to the total number of charmonia pairs. {\it Bottom:} A qualitative conjecture of dissociation and re-combination of \jpsi\ as a function of energy density. The probability is normalized per initial \jpsi\ particle (see \cite{Satz:2006kba}).}
\label{charmonia-suppression}
\end{center}
\end{figure}

Quarkonia (bound states of heavy quark and anti-quark) play historically an important role in establishing the existence and properties of QGP \cite{Matsui:1986dk}; however, the complete description of their production in heavy-ion collisions remains a challenge to theory. Charmonia states for instance are an attractive probe of the strongly coupled system of quarks and gluons as their sizes can be smaller than light hadrons (down to a few tenths of a fm) and having large binding energies ($>500$ MeV). Thus, early on it was suggested that the quarkonia states can be used as a {\it thermometer} of the QGP \cite{Matsui:1986dk,Mocsy:2013syh}. Various states would dissociate within the plasma depending on the screening potential of the QGP. The so-called sequential suppression was predicted: on average the heavy-ion collisions would produce fewer quarkonia states as compared to the extrapolation from the proton-proton collisions, and the states strongly bound would melt at higher temperatures than the less bound states. The color screening would dissociate charmonium in QGP such that first the higher excited states (2S), (1P), then the ground state (1S) would break up (see Fig. \ref{charmonia-suppression} showing evolution of \jpsi\ production probability as a function of energy density). The first predictions pointed that the \jpsi\ (1S) would not survive temperatures of about 2 times $T_{c}$, the critical temperature needed for creating the QGP. At present according to lattice QCD calculation this number is estimated between 1.2 and 1.9 $T_{c}$ depending on the details of the lattice calculations. On the other hand, a compensating scenario may take place. Namely, 1) all primary charmonia dissociated at high collision energy contribute to the abundance of charm quarks; and 2) charm and anti-charm quarks can equilibrate with the rest of the plasma and a substantial amount of the $c\bar{c}$ pairs may survive until hadronisation when by statistical combination charmonia states in addition to the open charm hadrons can be created (see for example Ref. \cite{Andronic:2018vqh}). Such scenario is presented in Fig. \ref{charmonia-suppression} in terms of production probability of \jpsi\ as a function of the energy density.

The extraction of the genuine effects of hot QGP on the charmonia states is complicated by the CNM effects. These can be of two types. One related to the modifications of the parton distribution functions within nuclei (the so-called shadowing and anti-shadowing) that may have a direct impact on the observed number of produced quarkonia. The second effect originates from the final probability for energy loss of quarkonia within the nuclei. This final-state energy loss in the cold nuclear matter may also directly influence the observed \pt\ spectrum of quarkonia states. Both of these effects are present in the case of AA collisions. An experimental access to these effects can be provided by the collisions of protons and nuclei (see \cite{Abelev:2013yxa,Acharya:2018kxc} and references therein).

\begin{figure}[htb]
\begin{center}
\includegraphics[width=0.35\textwidth]{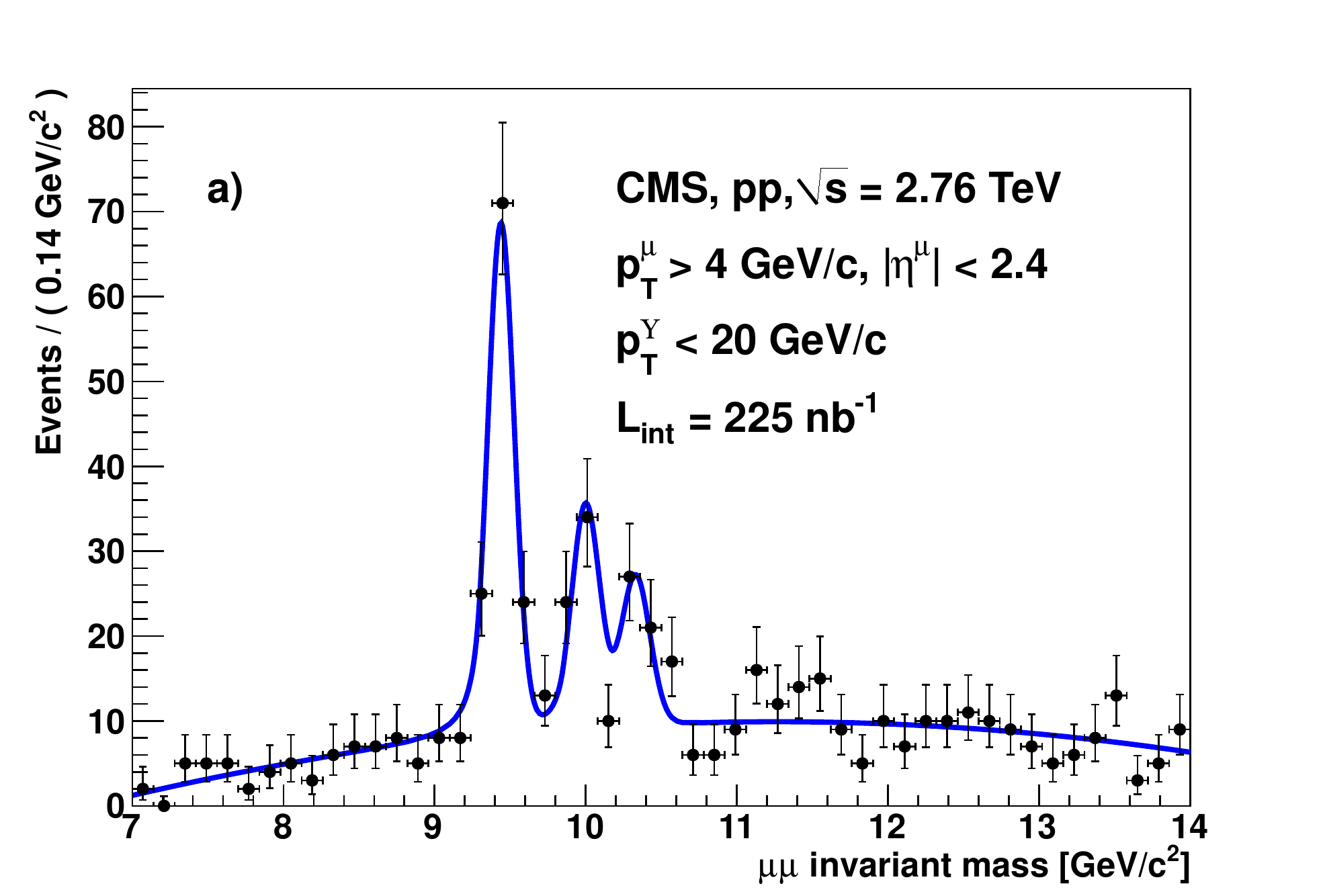}
\includegraphics[width=0.35\textwidth]{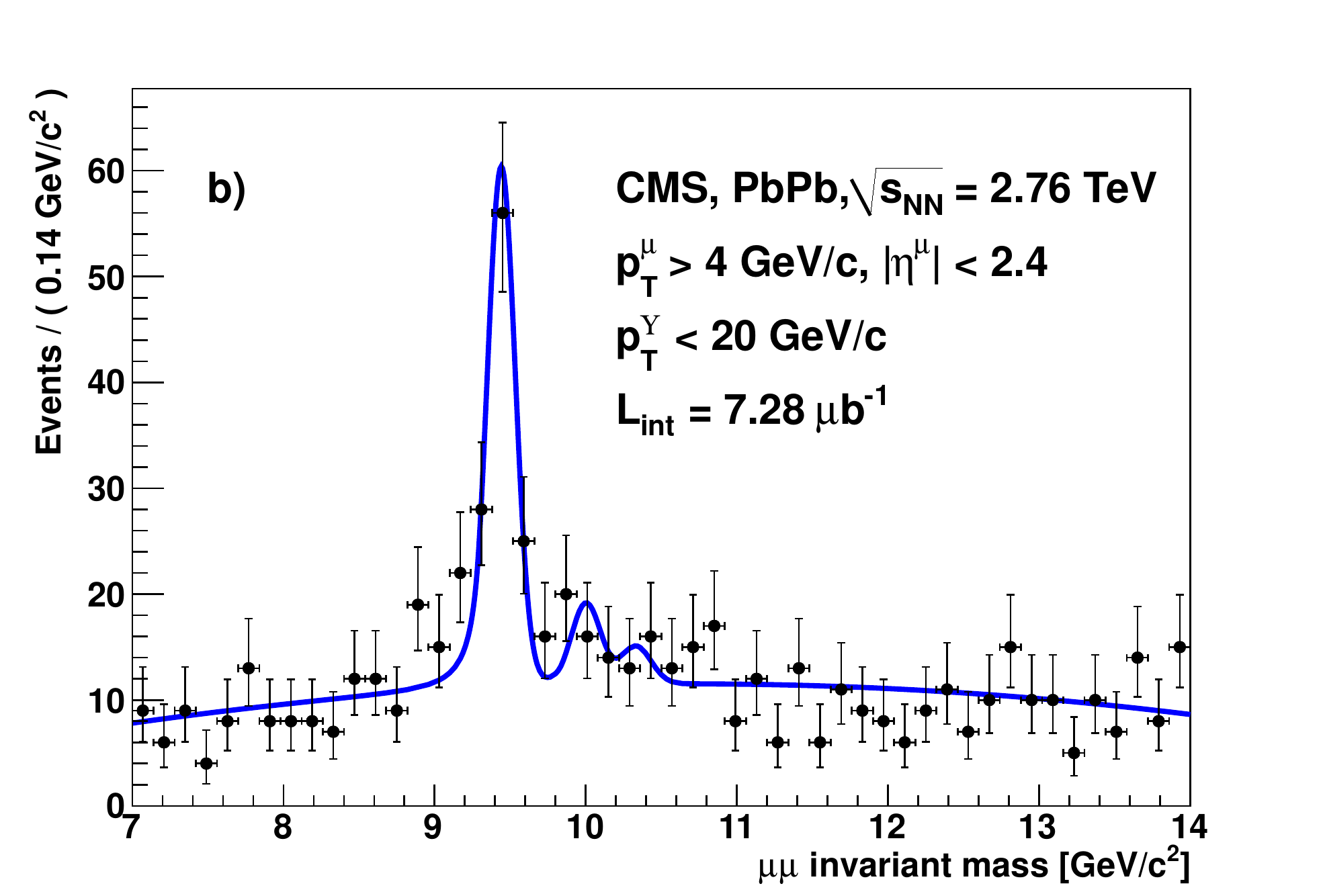}
\caption{Dimuon invariant-mass distributions from the \pp\ (a) and PbPb (b) data at $\sqrt{s_{\rm NN}} = 2.76$~TeV. The same reconstruction algorithm and analysis criteria are applied to both data sets, including a transverse momentum requirement on single muons of $p_{\rm T}^\mu >$~4~GeV/$c$. The solid lines show the result of the fit described in the text. Figure from \cite{Chatrchyan:2011pe}.}
\label{Mass_PbPb_nofit}
\end{center}
\end{figure}

One of the most striking experimental evidence for the suppression of the quarkonia states and its relation to their respective binding energy is shown in Fig. \ref{Mass_PbPb_nofit}. The invariant-mass distribution of di-muons was reconstructed in \pp\ and \PbPb\ collisions by the CMS collaboration. The plots focus on the mass region of the $b\bar{b}$ states. It is evident that the higher states of 2S and 3S are no longer resolvable in the \PbPb\ case. A very different case than in \pp\ collisions where the three peaks of the $Y$ states are visible.

\begin{figure}[tb]
\begin{center}
\includegraphics[width=0.5\textwidth]{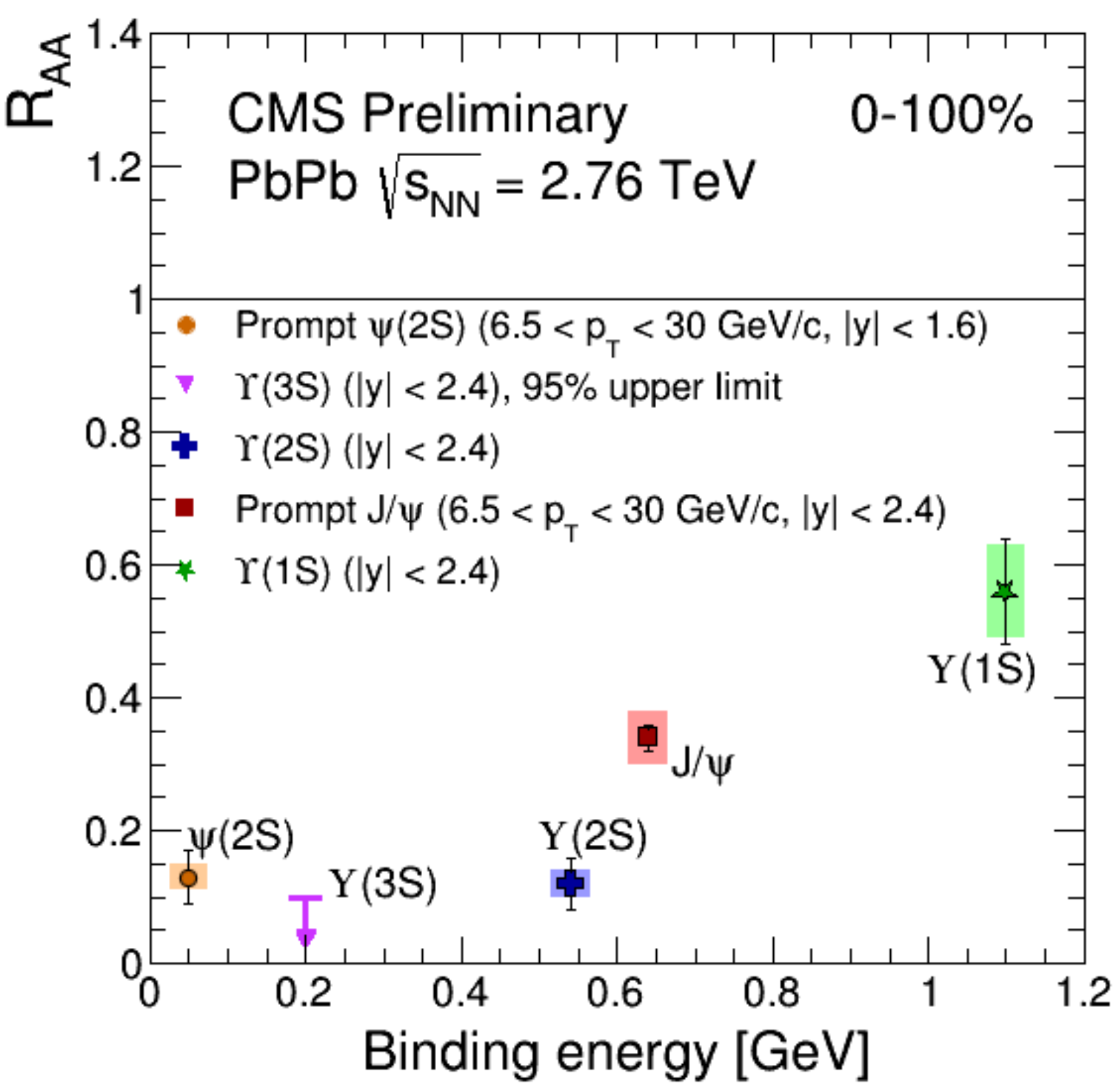}
\caption{Nuclear modification factor demonstrating the suppression of quarkonia states as a function of their binding energy. Figure based on \cite{Chatrchyan:2012lxa}.}
\label{CMS-seq-melting}
\end{center}
\end{figure}

The measurements of the modification factors for \jpsi\ and $Y$ states can be ordered with respect to their binding energy. Figure \ref{CMS-seq-melting} shows clearly the suppression pattern that is consistent with the expectation for the ``melting'' of the states according to strength of the binding. The states with small binding energy ($\psi$(2S), $Y$(3S) and $Y$(2S)) are strongly suppressed. While the \jpsi\ and $Y$(1S) show much larger \RAA.
Moreover, a clear evidence for heavy-quark thermalization and flow within the QGP medium is provided by measurements of elliptic flow $v_{2} > 0$ of \jpsi\ \cite{Acharya:2017tgv}. For more in-depth discussion of heavy-flavor and quarkonia in \pp\ and AA collisions see Ref. \cite{Andronic:2015wma}. For recent results on \jpsi\ production in Xe-Xe collisions and discussion of earlier results on the suppression see Ref. \cite{Acharya:2018jvc}.

\subsubsection{Density and Transport Coefficient}

More informtion on the transport properties and internal structure of the QGP can be obtained via measurements of interaction of the colored {\it hard probes} with such a medium.
The idea here is quite simple: similar to X-ray scanning of an unknown material, highly energetic (hard) partons that interact with the medium are used.
The first consideration of using particle jets resulting from the decay of highly virtual partons to probe the medium dates back to 1980's \cite{Bjorken:1982tu}.
For a more recent but pre-LHC review see \cite{Wiedemann:2009sh}.
Nowadays, at RHIC and even more so at the LHC energies, partonic interactions with large momentum transfer are abundant, and the hadronic remnants of high-energy scattered partons ({\it jets}) become experimentally accessible.
These partons while propagating through the dense matter produced in the collision may interact with the medium and loose a portion of their energy ({\it medium-induced energy loss}).
A parameter that is used to characterize the medium is the so-called jet transport coefficient $\hat{q} \propto \mu^{2} / \lambda$, where $\mu$ is the momentum transfer between the medium and the parton, and $\lambda$ is the mean free path within the medium.
The $\hat{q}$ is directly proportional to the density of the medium (through the dependence on the mean free path).
The energy loss of a parton within a medium of a given density and temperature can be related to $\hat{q}$ such that $\dedx \propto \alpha_{s} \hat{q} L^{2}$, where $L$ is the path length that a parton traversed through the medium.

\onecolumngrid

\begin{figure}[htb]
\begin{center}
\includegraphics[width=0.4\textwidth]{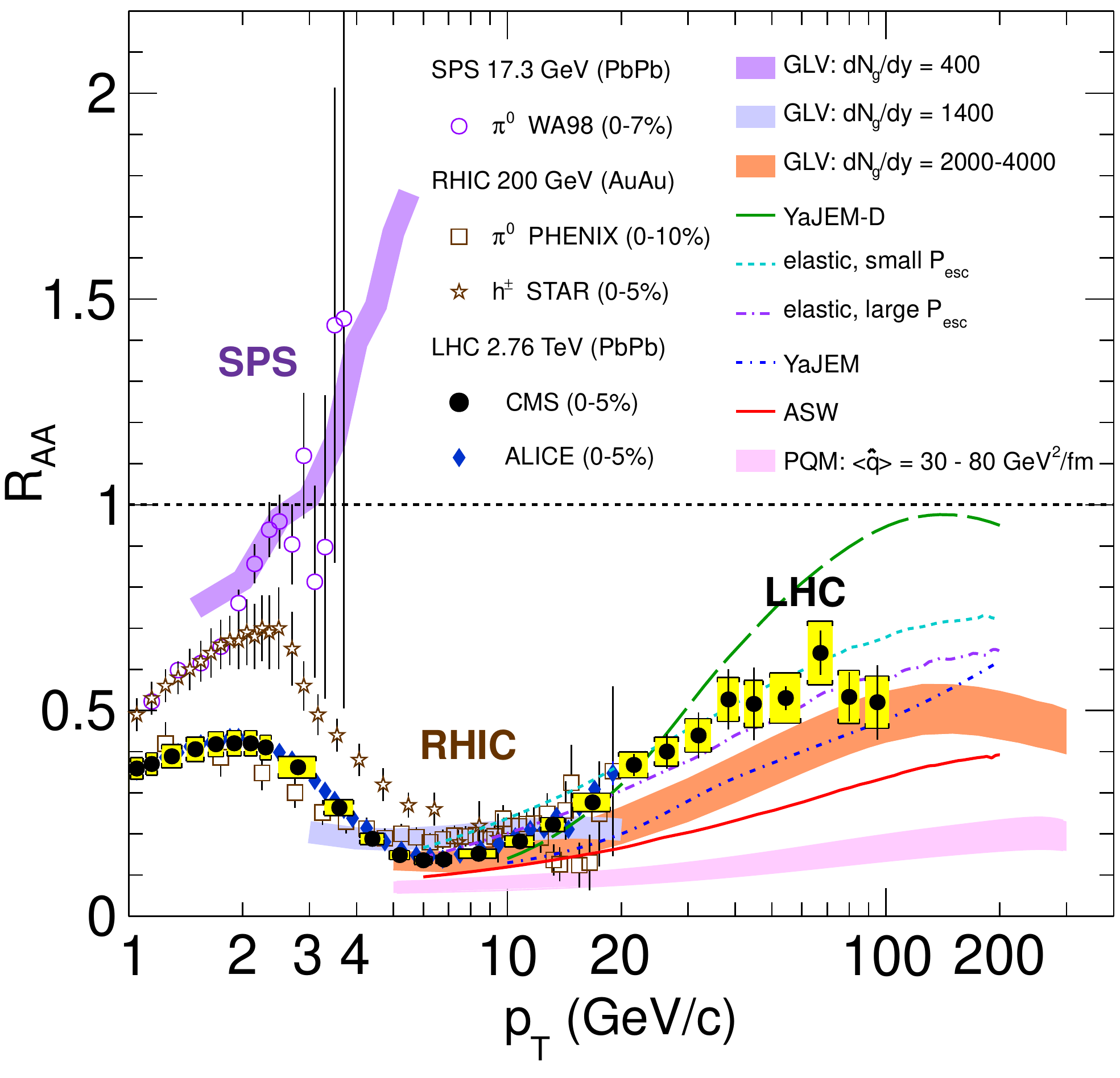}
\includegraphics[width=0.52\textwidth]{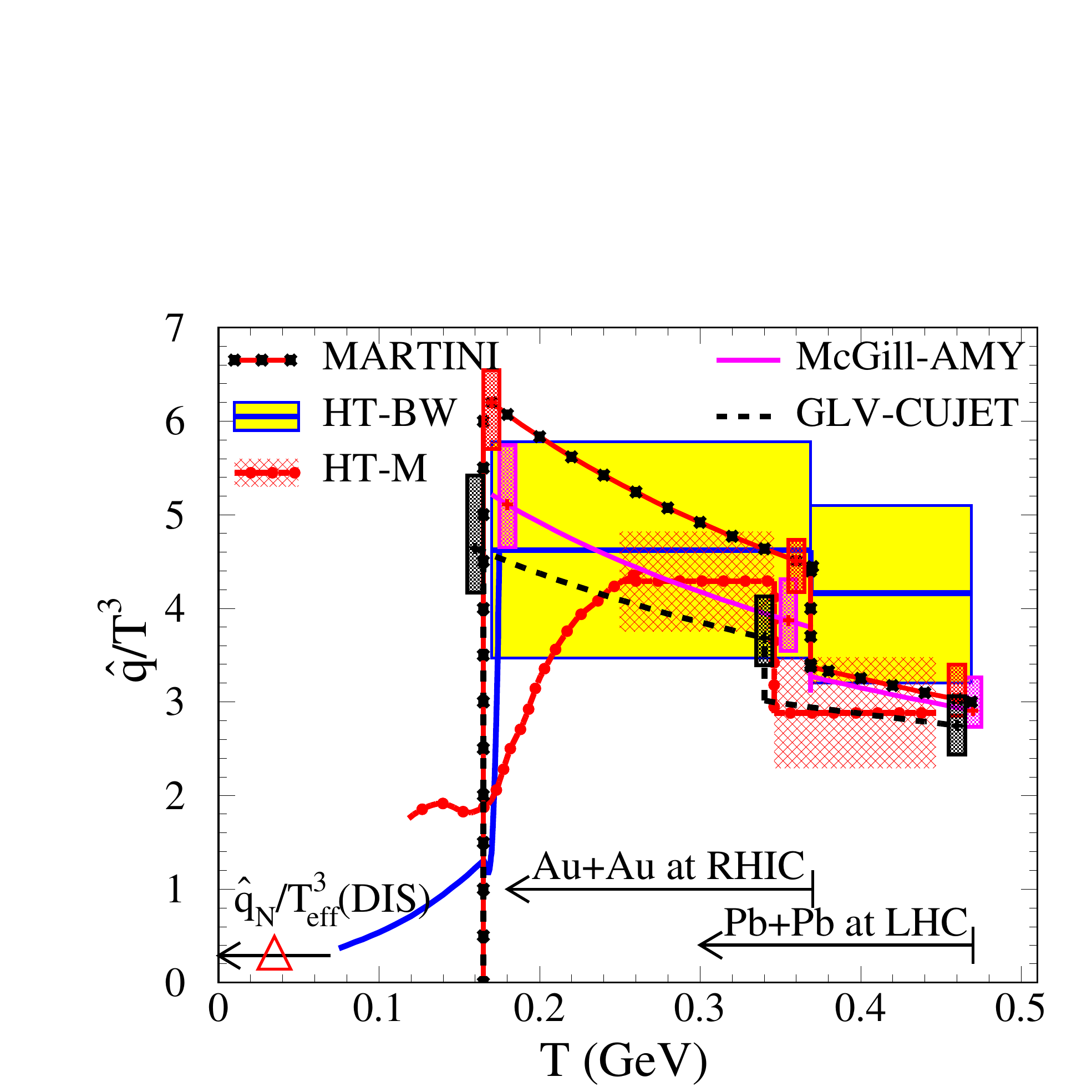}
\caption{{\it Left:} \RAA\ for charged hadrons measured at the SPS, RHIC and the LHC. The experimental results are compared to theoretical calculations (see \cite{CMS:2012aa} and references therein for details).
{\it Right:} The assumed temperature dependence of the scaled jet transport parameter $\hat q/T^3$ in different jet quenching models for an initial quark jet with energy $E=10$ GeV. Values of $\hat q$ were constrained by experimental data on hadron suppression $R_{AA}$ at both RHIC and the LHC.
The arrows indicate the range of temperatures at the center of the most central AA collisions. The triangle indicates the value of $\hat q_N/T^3_{\rm eff}$ in cold nuclei from deep inelastic scattering (DIS) experiments. Figure from \cite{Burke:2013yra}.}
\label{1202.2554/arxiv.org/abs/1202.2554/img/raa_compiled_QM11_square_hi2011.pdf}
\label{fig:qhat}
\end{center}
\end{figure}
\twocolumngrid

Historically, full jet reconstruction in the presence of large backgrounds in heavy-ion collisions was thought to be impossible. This view predominated until recent developments in background subtraction techniques. Therefore, initially the experiments at RHIC  utilized the single particle measurements, which approximated jets via leading hadron observables. STAR and PHENIX at RHIC have measured a strong depletion of high-\pT\ hadrons created in central heavy-ion collisions as compared to the expected yields derived from proton-proton measurements.

Assuming the scaling of hard processes with the number of independent binary nucleon-nucleon collisions one can define the nuclear modification factor \RAA\ as a ratio of hadron yields measured in heavy-ion collisions to expected yields obtained by superposition of independent nucleon-nucleon inelastic collisions:

\begin{align} 
\RAA(\pt) = \frac{{\rm d}N_{\rm h}^{\rm AA}\pt/{\rm d}\pt}{\langle N_{\rm bin} \rangle {\rm d} N^{\pp}_{\rm h}(\pt)/{\rm d}\pt}
\end{align}

The measured \RAA\ (see Fig. \ref{1202.2554/arxiv.org/abs/1202.2554/img/raa_compiled_QM11_square_hi2011.pdf}) indicates a large deficit (of about factor 5) of high-pT hadrons in \AuAu\ collisions as compared to p-p. At the LHC the \RAA\ grows to up to 0.5 at 100 \gevc. Such loss of hadrons and no sign of deficit for the color neutral objects (direct photons, $W$ and $Z$ bosons recently measured at the LHC - $\RAA \sim 1$  \cite{Chatrchyan:2012vq,Chatrchyan:2012nt,Chatrchyan:2011ua}), as well as no sign of nuclear effects at high-\pt\ in \pPb\ collisions at the LHC \cite{ALICE:2012mj} supports the argument of substantial interactions of high-energy partons within a hot and dense colored medium created in heavy-ion collisions.

Figure \ref{fig:qhat} (right panel) summarizes the current status of characterizing the properties of the medium using the jet transport parameter $\hat{q}/T^{3}$. The value of the parameter has been extracted by fitting the experimental data of single hadron \RAA\ and di-hadron at RHIC and the LHC, for a number of jet quenching models. In addition, the figure illustrates the evolution of the parameter with temperature (the ranges for RHIC and the LHC are indicated). For more details see Ref. \cite{Burke:2013yra}.

\begin{figure}[htb]
\begin{center}
\includegraphics[width=0.45\textwidth]{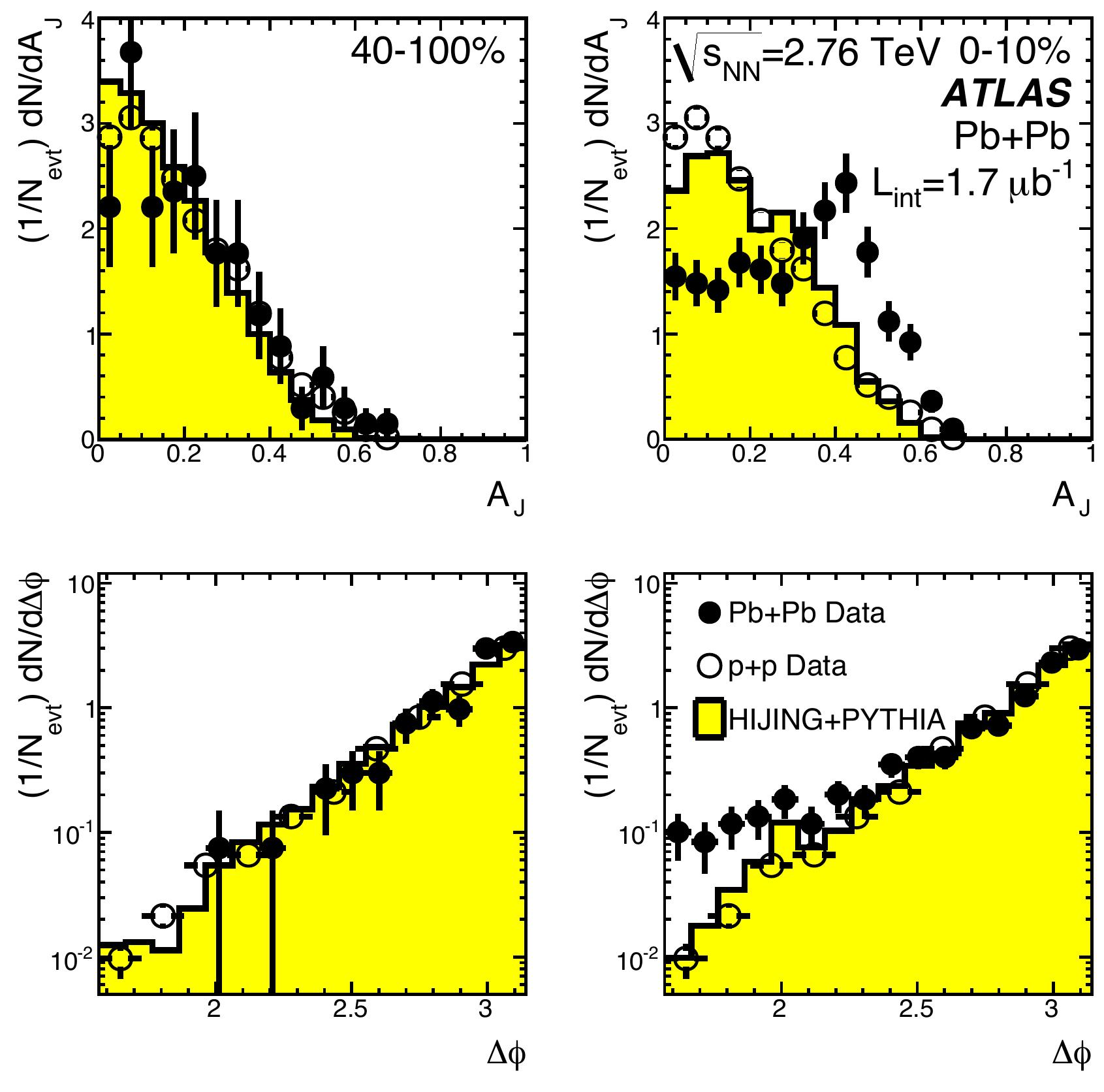}
\caption{(top) Dijet asymmetry distributions for data (points) and unquenched HIJING with superimposed PYTHIA dijets (solid yellow histograms), as a function of collision centrality (left to right from peripheral to central events). Proton-proton
data from \sqrts = 7 TeV, analyzed with the same jet selection, is shown as open circles. (bottom) Distribution of $\Delta \varphi$, the azimuthal angle between the two jets, for data and HIJING+PYTHIA, also as a function of centrality. Figure from \cite{Aad:2010bu}.}
\label{1011.6182./arxiv.org/abs/1011.6182/img/final_4x2_23_newpp.pdf}
\end{center}
\end{figure}

The first measurement of jet quenching with fully reconstructed jets in heavy-ion collisions focused on the balance between the highest transverse energy pair of jets in events where the two jets have an azimuthal angle separation, $\Delta \varphi = |\varphi_{1}-\varphi_{2}| > \pi/2$. ATLAS Collaboration has analyzed \PbPb\ collisions at \sqrtsnn{2.76} and extracted the dijet asymmetry $A_{j}$ defined as:
\begin{align} 
A_{j} = \frac{E_{\rm T1} - E_{\rm T2}}{E_{\rm T1} + E_{\rm T2}},
\end{align}
where the first jet is required to have a transverse energy $E_{\rm T1} > 100$~GeV, and the second jet is the highest transverse energy jet in the opposite hemisphere with $E_{\rm T2} > 25$~GeV. The jets were reconstructed using the anti-\kt algorithm with $R=0.4$. The results shown in Fig. \ref{1011.6182./arxiv.org/abs/1011.6182/img/final_4x2_23_newpp.pdf} have two distinct features:

\begin{itemize}
\item The asymmetry in peripheral events is consistent with the \pp\ data while for central events it shows strong modifications: a depletion for small $A_{j}$ and a maximum at about 0.4. This is a clear indication that the recoiling jet loses a significant fraction of its energy.
\item The accoplanarity of the dijet system (demonstrated by the $\Delta \varphi$ distribution) remains unchanged with respect to \pp\ collisions. This is observed even for the most central collisions where the quenching effects are expected to be the strongest.
\end{itemize}

These two observations provide a wealth of information about the nature of jet quenching beyond the single hadron and jet \RAA. The modification of the $A_{j}$ confirmed the conclusion that the medium induced radiation is transported outside of the jet cone and it is consistent with the $\RAA < 1$ (or $\RCP < 1$). On the other hand, the lack of medium induced acoplanarity provides strong constraints to the theoretical considerations of the quenching process: although jets loose significant amount of energy, their original direction remains unmodified.

The abundance of high-\pt\ probes at the LHC and RHIC has provided a flare of new results on jet spectrum modifications but also on jet substructure modifications.
Nowadays, it is a remarkably quickly evolving subfield.
For a recent overviews see Ref. \cite{Cunqueiro:2017bld,Mehtar-Tani:2016aco} and references therein.

Heavy-quarks are produced early in heavy-ion collisions and their abundances do not change through the evolution of the collision system.
They are produced in hard collisions (because of their large mass) and their production can be calculated within the pQCD framework.
This makes them excellent auto-generated probes of the medium.
However, the central point of the considerations of heavy-quark energy loss is that the gluon radiation from a highly energetic parton traversing the quark-gluon plasma shows a characteristic hierarchical dependence on the color charge and mass of the parton projectile: $\Delta E _{\rm gluon} > \Delta E _{\rm light~quark} > \Delta E _{\rm heavy~quark}$.
Here, the first inequality follows from the larger color charge of partons in the adjoint representation.
The second inequality is due to the dead cone effect, which suppresses radiation of massive particles in the vacuum and in the medium.
Indeed, theoretical calculation show that the energy loss of a quark is inverse proportional to its mass ($\Delta E \sim 1/M$) and provides a handle on the longitudinal diffusion coefficient \cite{Majumder:2008zg,Armesto:2003jh} of the QGP.
The recent measurements at RHIC and the LHC \cite{Dong:2017dws} show that indeed heavy-flavor quarks loose substantial energy within the medium and flow within the bulk of QGP \cite{Rapp:2018qla}.
Moreover, measurements of B-mesons and non-prompt \jpsi\ provide instight into the parton mass dependence of the energy loss.
For a projections of heavy-flavor measurements in LHC Run-3 and references to current data see \cite{CMS-PAS-FTR-17-002,Aiola:2017uym}.

\section{\label{sec:summary}Summary}

This write up provides only a glimpse at the physics of the hot and dense QGP.
The heavy-ion collision field is continuing its mission to understand the nature of matter at extreme conditions, both at high densities and high temperatures.
In the upcoming years RHIC will continue exploration of the phase diagram in the search of critical point with the Beam Energy Scan and LHC will continue providing most interesting data from the high energy regime to study the inner working of the hot plasma.

\section*{Acknowledgements}
I would like to kindly thank to the organizers of the XIV International Workshop in Hadron Physics, particularly Prof. Menezes, Prof. Benghi, and Dr. Oliveira, for their kindness, hospitality, and the great effort in organizing an excellent workshop.
The financial support of the The Brazilian National Council for Scientific and Technological Development most appreciated.
This work was supported in part by the U.S. Department of Energy, Office of Science, Office of Nuclear Physics, under contract DE-AC02-05CH11231.

\bibliography{main}

\end{document}